\documentclass[preprint,showpacs,preprintnumbers,amsmath,amssymb,superscriptaddress,aps,floatfix,prc,nofootinbib]{revtex4-1}

\usepackage{longtable}
\usepackage{graphicx}
\usepackage{dcolumn}
\usepackage{bm}
\usepackage[dvipsnames,usenames]{color}
\usepackage{subfigure}
\usepackage{multirow}
\usepackage{amssymb}
\usepackage{mdwlist}
\usepackage{lineno}

\def\today{\space\number\day\space\ifcase\month\or January\or February\or
    March\or April\or May\or June\or July\or August\or September\or October\or
    November\or December\fi\space\number\year}

\newcommand{\iso}[2]{$^{#1}$#2}

\begin{document}

\setpagewiselinenumbers
\linenumbers

\preprint{Draft 0.2}

\title{Comparison of Lithium Gadolinium Borate Crystal Shards in Scintillating and Nonscintillating Plastic Matrices}

\author{K. Kazkaz\footnote{Corresponding author, kareem@llnl.gov}}\affiliation{Lawrence Livermore National Laboratory, 7000 East Ave., Livermore CA 94551}
\author{N.\,S. Bowden}\affiliation{Lawrence Livermore National Laboratory, 7000 East Ave., Livermore CA 94551}
\author{M. Pedretti}\affiliation{Lawrence Livermore National Laboratory, 7000 East Ave., Livermore CA 94551}

\date{\today}

\begin{abstract}
We present a method for detecting neutrons using scintillating lithium gadolinium borate crystal shards in a plastic matrix while maintaining high gamma rejection. We have procured two cylindrical detectors, 5''$\times$5'', containing 1\% crystal by mass. Crystal shards have a typical dimension of 1~mm. One detector was made with scintillating plastic, and one with nonscintillating plastic. Pulse shape analysis was used to reject gamma ray backgrounds. The scintillating detector was measured to have an intrinsic fast fission neutron efficiency of 0.4\% and a gamma sensitivity of less than $2.3 \times 10^{-9}$, while the nonscintillating detector had a neutron efficiency of 0.7\% and gamma sensitivity of $(4.75 \pm 3.94) \times 10^{-9}$. We determine that increasing the neutron detection efficiency by a factor of 2 will make the detector competitive with moderated \iso{3}{He} tubes, and we discuss several simple and straightforward methods for obtaining or surpassing such an improvement. We end with a discussion of possible applications, both for the scintillating-plastic and nonscintillating-plastic detectors.
\end{abstract}

\maketitle

%
%
\section{Introduction}
\label{s:Intro}

Neutron detectors are useful in a variety of situations, from monitoring nuclear power plants~\cite{Knoll2010}, to imaging~\cite{Anderson2009}, to fissile material localization~\cite{Bowden2010}, and so on. Depending on the specifics of the detector, they can be optimized for spectroscopy, directionality, scalar counting with or without timestamps, or some combination of the three.

One detector technology that can be used for either counting or spectroscopy utilizes embedded inorganic scintillator crystals in a plastic matrix. This method was first proposed almost 10~years ago by Czirr {\it et al.} for the purposes of capture-gated neutron spectroscopy~\cite{Czirr2002}. In that work, the crystal Czirr {\it et al.} used was lithium gadolinium borate (LGB)~\cite{Czirr1999}, with the chemical composition Li$_6$Gd(BO$_3$)$_3$, where the lithium and/or boron was enriched in the high-capture-cross-section isotopes $^6$Li and $^{10}$B for various measurements. Capture-gated spectroscopy works by looking for a double light pulse, the first pulse coming from neutron recoils in the organic scintillator, and the second coming from the capture of that same neutron. The the case of LGB/plastic composite scintillators, interactions were differentiated by exploiting the very different scintillation decays times: a few ns for plastic, and 270~ns for the crystal.

While it was possible for neutrons to recoil in the crystal rather than the plastic, this was unlikely in Ref.~\cite{Czirr2002} because the detector was 90\% plastic by mass. In a similar vein, it was also possible for the neutrons to capture on a proton in the plastic rather than the crystal, although this effect was subdominant because of the high macroscopic neutron capture cross-section of the crystal. A simple, though not absolutely precise, correlation could therefore be made, where short pulses were caused by neutron recoils, and long pulses caused by neutron captures. This ansatz extends to gamma recoils being typified by short pulses, similar to neutron recoils.

Further particle discrimination is available using energy discrimination. For example, a neutron capture on \iso{6}{Li} liberates 4.78~MeV of kinetic energy. After scintillation quenching, this energy deposition in LGB produces the same number of optical photons as a 2.2~MeV electron~\cite{Czirr2002}. This electron-equivalent energy is above most background radiation, with the ubiquitous 2.6~MeV gamma ray from \iso{208}{Tl} being a notable exception. Though the number of scintillation photons may be the same, a 2.2-MeV electron has a path length of approximately 4.4~mm~\cite{EStar2011}, so if the LGB crystals have a typical dimension of, for example, 1~mm, a 2.2-MeV electron will create only $\sim$500~keV of visible energy from the LGB shard. By comparison, the recoiling tritium resulting from a capture on \iso{6}{Li} has a range of only $\sim$30~$\mu$m~\cite{Ziegler2008}. Combining both energy analysis and pulse shape analysis is therefore capable of providing very strong discrimination between neutron captures and neutron/gamma recoils, as discussed below in Section~\ref{s:Experiment}.

Additional work has been performed on LGB-embedded, scintillating-plastic detectors. Menaa {\it et al.} have evaluated 2''$\times$2'' cylindrical LGB / scintillating-plastic detectors for possible use as a hand-held neutron spectometer~\cite{Menaa2009}. Flaska, Pozzi, and Czirr have studied a 5" diameter, 4" long detector from the standpoint of pulse shape discrimination~\cite{Flaska2008}. Nelson and Bowden have studied such detectors to determine their suitability for anti-neutrino detection~\cite{Nelson2011}.

The current work builds on the previous efforts in a number of ways. In Section~\ref{ss:PSA} we discuss methods to improve the pulse shape discrimination using additional information obtained from a digitized signal. In Section~\ref{s:Optimizations} we use Monte Carlo studies in an effort to increase neutron sensitivity while keeping gamma rejection at acceptable levels. Finally, in Section~\ref{s:Improvements} we briefly cover additional methods to improve detector performance.

%
%
\section{The Experimental Program}
\label{s:Experiment}

We began our studies with the same detector, data acquisition, and analysis as outlined in Ref.~\cite{Nelson2011}. In that work, the detector, manufactured by MSI/Photogenics, was a cylindrical, 5''$\times$5'' detector and 1\% by mass LGB. The lithium and boron in the crystals were enriched to $\sim$95\% \iso{6}{Li} and \iso{10}{B}, while the gadolinium isotopes were of natural abundances. The relevant nuclear reactions were therefore

\begin{equation}
\begin{array}{lllll}
^6\mbox{Li} + \mbox{n} & \rightarrow & ^3\mbox{H}~(2.75 \mbox{~MeV}) +~^4\mbox{He}~(2.05 \mbox{~MeV}) & ~ \\
^{10}\mbox{B} + \mbox{n} & \rightarrow & ^7\mbox{Li}~(1.0 \mbox{~MeV}) +~^4\mbox{He}~(1.8 \mbox{~MeV}) & ~\mbox{BR = 7\%} \\
~ & \rightarrow & ^{*7}\mbox{Li}~(0.83 \mbox{~MeV}) +~^4\mbox{He}~(1.47 \mbox{~MeV}) & ~\mbox{BR = 93\%} \\
~ & ~ & \multicolumn{2}{l}{^{*7}\mbox{Li} \rightarrow ~^7\mbox{Li} + \gamma~(0.48 \mbox{~MeV})} \\
^{\mbox{x}}\mbox{Gd} + \mbox{n} & \rightarrow &^{\mbox{x+1}}\mbox{Gd} + \gamma~(\sim8~\mbox{MeV cascade})
\end{array}
\label{eq:LGBReactions}
\end{equation}

The LGB shards had a typical dimension of 1~mm, and the plastic used was EJ-290, a scintillating polyvinyltoluene from Eljen Technology, Inc. Two 5" Adit model B133D01W photomultiplier tubes, on either side of the detector, were connected to the data acquisition system. Attenuation in the light signal was taken into account using standard analysis techniques, as described in Eq.(1) in Ref.~\cite{Nelson2011}.

\begin{figure}[b!!!]
\centering
\includegraphics[width=8.5cm]{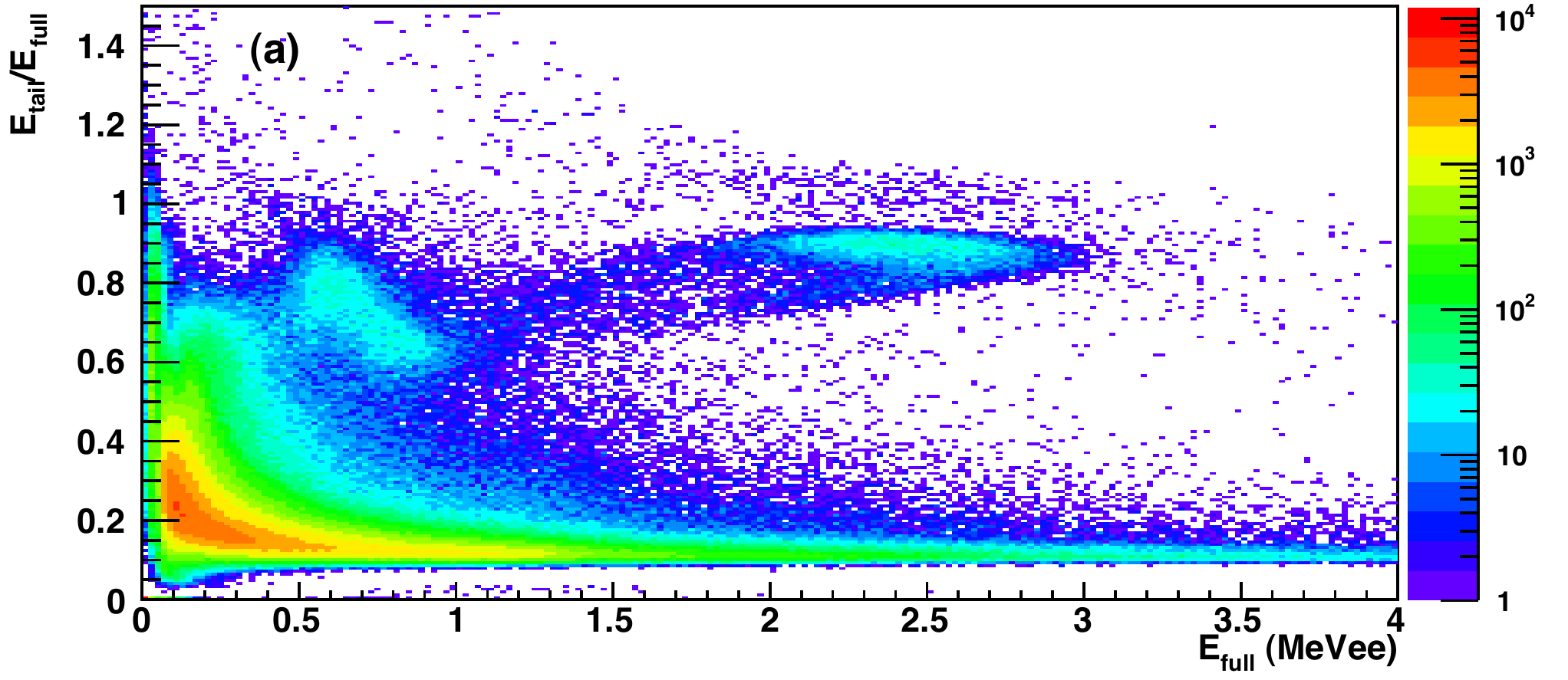}
\parbox{8.5cm}{\vspace{5pt}\caption{\small{Results from the scintillating plastic detector using an analog data acquisition system. This plot, along with an explanation of the pulse shape parameter, are from Ref.~\cite{Nelson2011}. The captures in \iso{6}{Li} are clearly distinguishable as the island between 2 and 3~MeVee and 0.8 to 1.0 on the vertical axis. A second island, much closer to the gamma background, is also visible between 0.5 and 1~MeVee and 0.6 to 0.8 on the vertical axis. This second island is associated with captures on \iso{10}{B}.}}
\label{fig:OldAnalogResults}}
\end{figure}

An analog data acquisition system was used to identify captures on \iso{6}{Li} by plotting the detector responses in a two-dimensional histogram, with the pulse shape parameter along the vertical axis and energy along the horizontal axis. Fig.~\ref{fig:OldAnalogResults} shows a clear signal island resulting from the capture of neutrons on \iso{6}{Li}.

With this plot as a starting basis, we undertook several approaches to increasing the fidelity of the neutron capture signal, while increasing gamma rejection. We then measured neutron efficiency and gamma rejection using both the scintillating-plastic and nonscintillating-plastic detectors. This work is detailed in the rest of this section.

\subsection{Data acquisition}
\label{ss:DAQ}

In Ref.~\cite{Nelson2011}, an analog data acquisition system was used, allowing for a relatively simple pulse shape analysis using a ratio of the integral of the tail of the pulse to the integral of the full pulse. This tail / full ratio is the vertical axis of Fig.~\ref{fig:OldAnalogResults}. Using this simple discrimination, there are several event topologies that can fake a neutron signal. For example, a low-energy electron recoil in the plastic may trigger the data acquisition system, and if an accidental high-energy electron recoil in the plastic falls within the tail gate time, the ratio of the tail integral to the / full integral may be close to 1, mimicking the tail / full ratio of a neutron capture on \iso{6}{Li}.

\begin{figure}[b!!!]
\centering
\includegraphics[width=8.5cm]{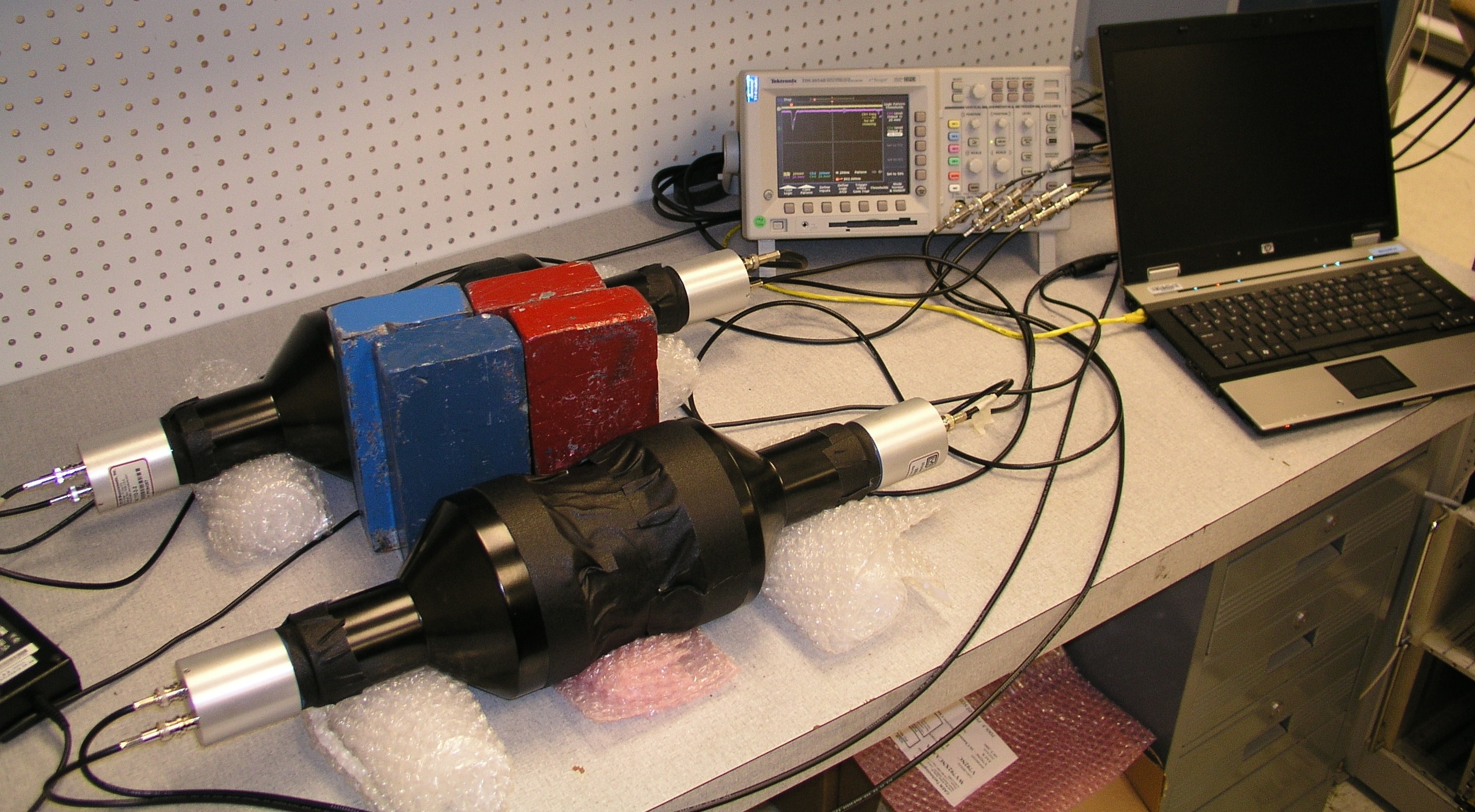}
\parbox{8.5cm}{\vspace{5pt}\caption{\small{A photograph of the detector setup used in this work. Each detector had two photomultiplier tubes, which were run into an oscilloscope for testing purposes, and then on to the data acquisition system (not visible in this photo). The lead bricks between the detectors was used to reduce gamma cross-talk between the detectors.}}
\label{fig:DualDetectorSetup}}
\end{figure}

To address this and other false neutron signals, we moved to a digital data acquisition system to provide finer control over the pulse shape analysis. We connected the photomultiplier tubes of both the scintillating-plastic and nonscintillating-plastic detectors to a Struck 3320 digitizer, sampling at 200 MHz and 12 bits. The interface to the VME crate was a Struck 3150 controller, and the data acquisition computer was running Linux. The data acquisition software was custom developed at Lawrence Livermore National Laboratory. Fig.~\ref{fig:DualDetectorSetup} shows a photograph of the detector setup.

\begin{figure}[b!!!]
\centering
\includegraphics[width=8.5cm]{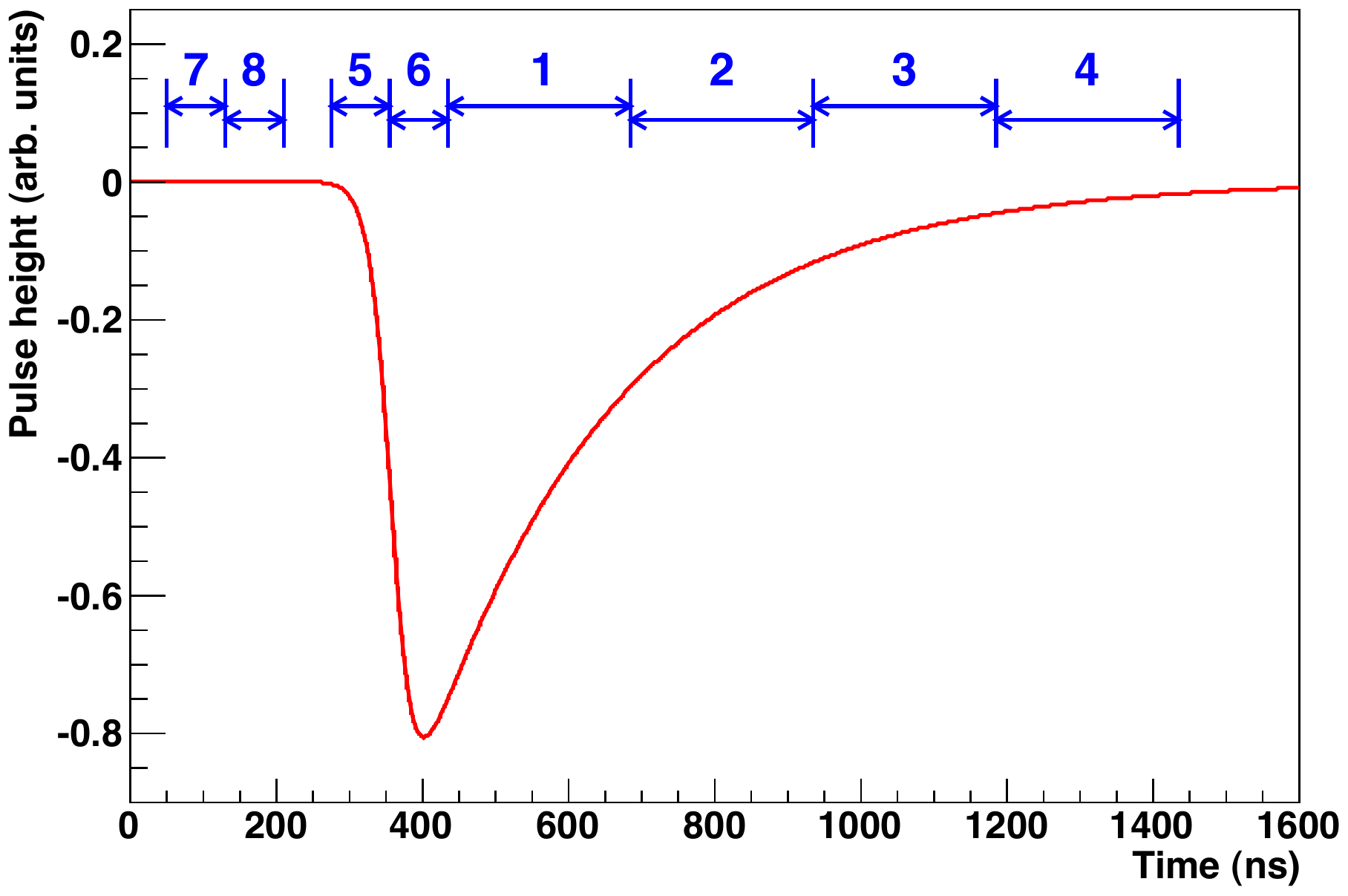}
\parbox{8.5cm}{\vspace{5pt}\caption{\small{Setup of the 8 integral gates on the Struck 3320. Gates 7 and 8 measure the baseline, gates 5 and 6 measure the rising edge, and gates 1-4 are used for pulse shape analysis. Summing gates 1-6 and subtracting the baseline contribution gives the full integral of the pulse. The curve is an idealized signal from the LGB crystal for clarity of explanation, and is not an experimental pulse.}}
\label{fig:GateSetup}}
\end{figure}

The Struck 3320 firmware used allows for defining eight integral gates, four of which can be up to 512 samples each, and four of which can be up to 16 samples each. At a sample rate of 200 MHz, this translates to maximum gate times of 2560 ~ns for each gate in the first group of four, and 80~ns for each gate in the second group of four. The actual gate lengths used in the analysis were 50 samples for gates 1-4 and 16 samples for gates 5-8. Note that the start times of the gates can be arbitrarily set, allowing not only for gaps between gates, but overlapping gates if desired. Fig.~\ref{fig:GateSetup} shows an idealized pulse from a neutron capture on \iso{6}{Li}, and how the gates were set up to subdivide the pulse. For every event, we used the baseline values as measured by gates 7 and 8 to obtain baseline-subtracted integrals on gates 1-6. This baseline subtraction of course took into account the differing gate widths.

The best possible pulse shape discrimination would have been possible if we recorded each entire pulse and ran pulse shape analysis on that, rather than relying on simple gate integrals. This would have introduced a great amount of computing overhead, though, as 400 points would have to be transferred to the DAQ computer and analyzed, rather than just 8. Because we used the gate integrals, we were able to maintain DAQ livetime above 99\%, even when the event rate was on the order of 5~kHz.

\subsection{Improving the pulse shape analysis}
\label{ss:PSA}

Once the digital data acquisition was set up, we re-created the analog pulse shape analysis by constructing a tail / full ratio:

\begin{equation}
\mbox{Analog-like pulse shape parameter} = \frac{\mbox{Integral of gates 1-4}}{\mbox{Integral of gates 1-6}}
\label{eq:AnalogPulseShapeParameter}
\end{equation}

\begin{figure}[t!!!]
\centering
\subfigure{\includegraphics[width=8.5cm] {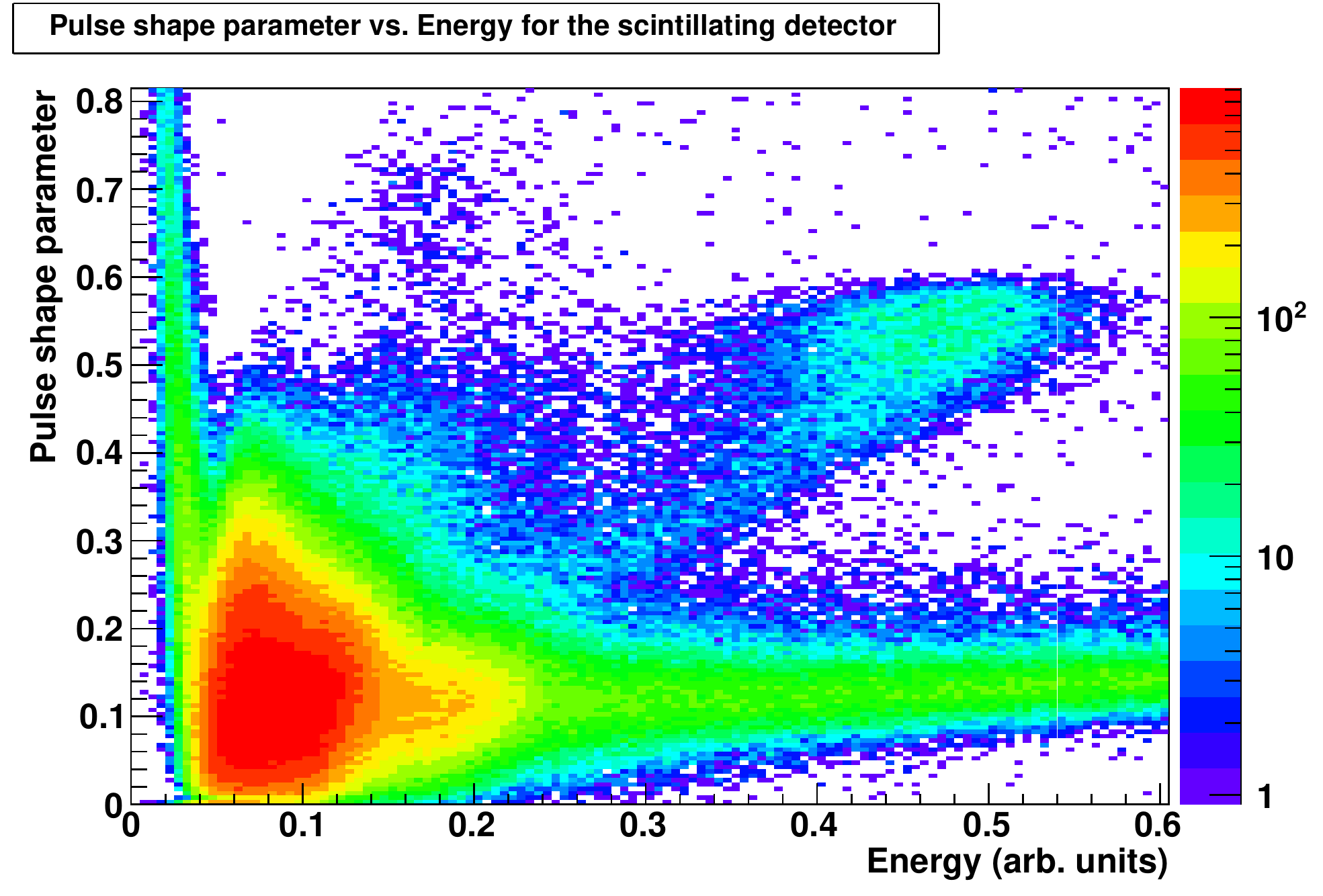}}
\subfigure{\includegraphics[width=8.5cm] {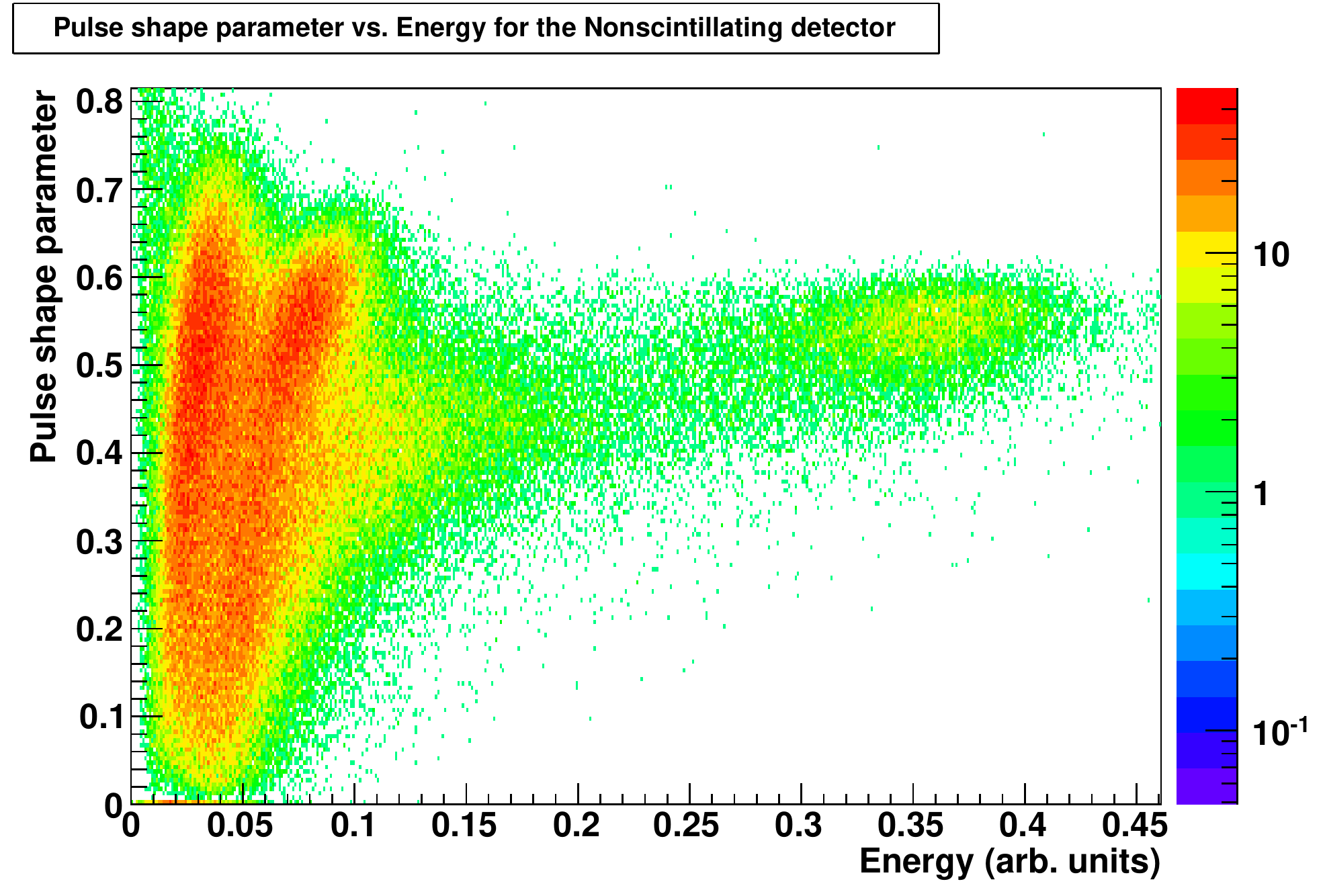}}
\parbox{8.5cm}{\vspace{5pt}\caption{\small{Analog-like data analysis of the scintillating-plastic (top) and nonscintillating-plastic (bottom) detectors. The tail gate was not identical to that used in the previous work, so the pulse shape parameter calculations will not be identical. Compare the top plot to Fig.~\ref{fig:OldAnalogResults}. The analog-like data analysis from this work differs from that shown in Fig.~\ref{fig:OldAnalogResults} because of differences in PMT response, integral gates, and threshold levels.}}
\label{fig:NewAnalogResults}}
\end{figure}

\noindent
The results of this analysis for both detectors is shown in Fig.~\ref{fig:NewAnalogResults}. The data in these plots were obtained using a bare, 75.4~kBq \iso{252}{Cf} source (8690 neutrons/second) suspended approximately 3~feet above the detectors. Data was accumulated over 20~hours, and the live time was over 99\%. For both detectors the \iso{6}{Li} island is clearly visible. The very broad gamma band between pulse shape parameter 0.1 and 0.2 in the scintillating-plastic detector is absent from the nonscintillating-plastic detector, while the amplitudes of the \iso{6}{Li} islands are of similar size.

\begin{figure}[t!!!]
\centering
\includegraphics[width=8.5cm]{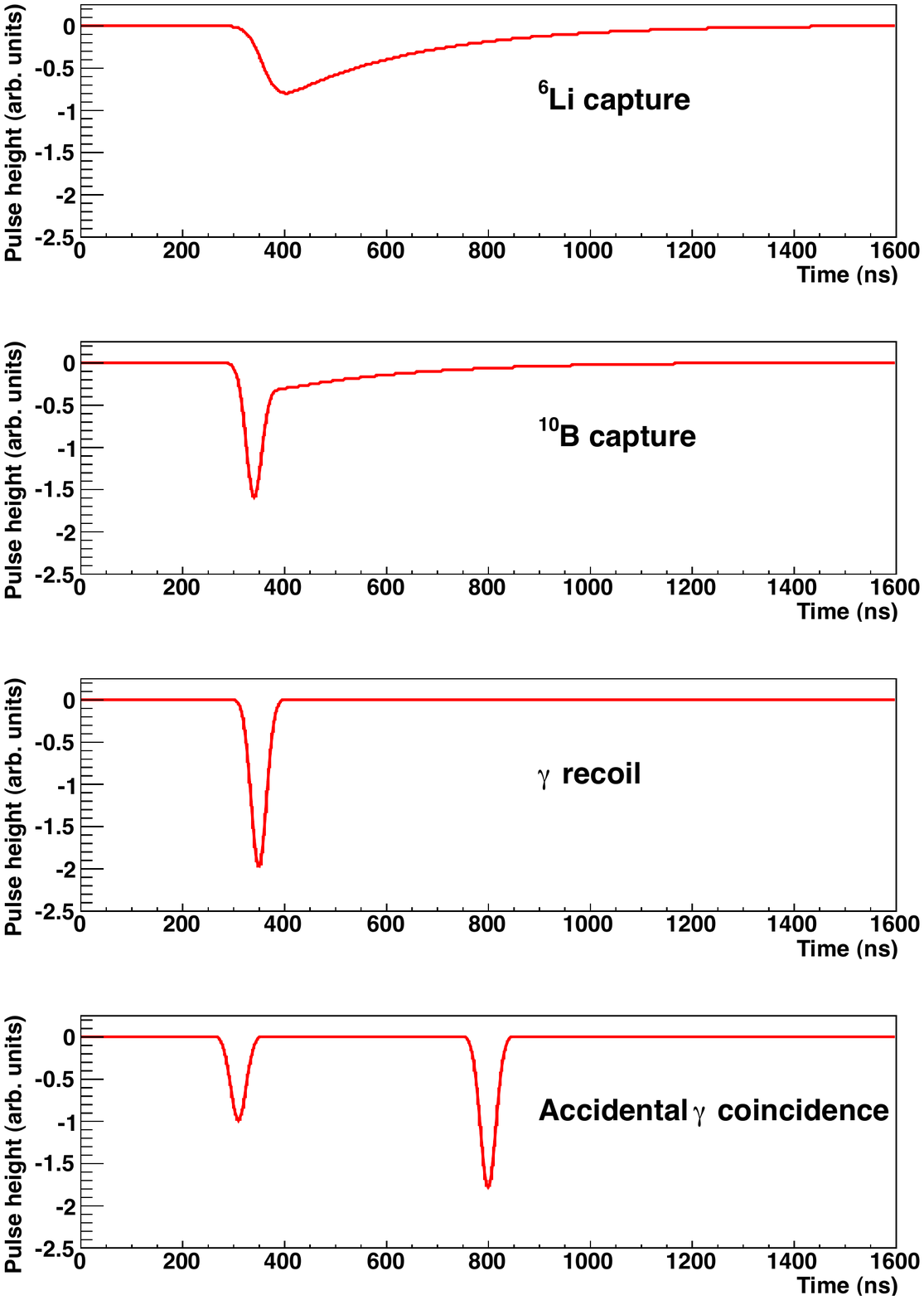}
\parbox{8.5cm}{\vspace{5pt}\caption{\small{Possible pulses from the scintillating-plastic LGB detector. These pulses are idealized pulses, and not from experimental data. See text for details.}}
\label{fig:ExamplePulses}}
\end{figure}

We have identified multiple event topologies that lead to differing effects when constructing the two-dimensional plots such as are shown in Fig.~\ref{fig:NewAnalogResults}. Fig.~\ref{fig:ExamplePulses} shows possible pulses from the scintillating-plastic LGB detector. The \iso{10}{B} capture shape results from the cascade gamma depositing energy in the plastic while the ion recoils remain within the LGB crystal. This pulse topology can also result if a neutron capture within an LGB shard results in ions escaping the crystal, such that energy is deposited in both the plastic and the crystal. Similarly, a gamma recoil can result in an electron depositing energy in both the plastic and a crystal shard.

\clearpage
Fig.~\ref{fig:ExamplePulses} also demonstrates the maximum data rate of these composite detectors. The Struck 3320 has data buffers that allow it to digitize a signal while transferring previously-accumulated data to a computer. This results in very low dead time, as discussed in Section~\ref{ss:Measurements}. Because each pulse is digitized for $\sim$1~$\mu$s, however, it is possible for multiple events to arrive within a single digitization time window. For example, the second gamma pulse in the bottom curve of Fig.~\ref{fig:ExamplePulses} could just as well be a neutron-capture event. If this were the case, the second pulse would be lost as part of event pile up. Because of the $\mu$s-scale digitization window, if the event rate were greater than $\sim$100 kHz, pile up would significantly increase.

Using the simple pulse shape analysis, an accidental gamma coincidence can create a background to the neutron capture signal. Referring to the bottom curve of Fig.~\ref{fig:ExamplePulses}, we note that the ``tail'' of the pulse, integrated from 500~ns to the end of the pulse, is a sizeable proportion of the full integral. This has the tendency to pull the pulse shape parameter closer to unity, and the event ends up encroaching on the neutron capture regions. We can avoid these kinds of backgrounds by making additional use of the gate integrals depicted in Fig.~\ref{fig:GateSetup}.

The approach we used to enhance the pulse shape discrimination was to use the ratios of

\begin{equation}
\begin{array}{c}
\left ( \mbox{Gate~} 5 + \mbox{Gate~} 6 \right )~/~\mbox{Gate~} 1\\
\mbox{Gate~} 1~/~\mbox{Gate~} 2 \\
\mbox{Gate~} 2~/~\mbox{Gate~} 3 \\
\mbox{Gate~} 3~/~\mbox{Gate~} 4
\end{array}
\label{eq:GateRatios}
\end{equation}
 
\begin{figure}[b!!!]
\centering
\includegraphics[width=8.5cm]{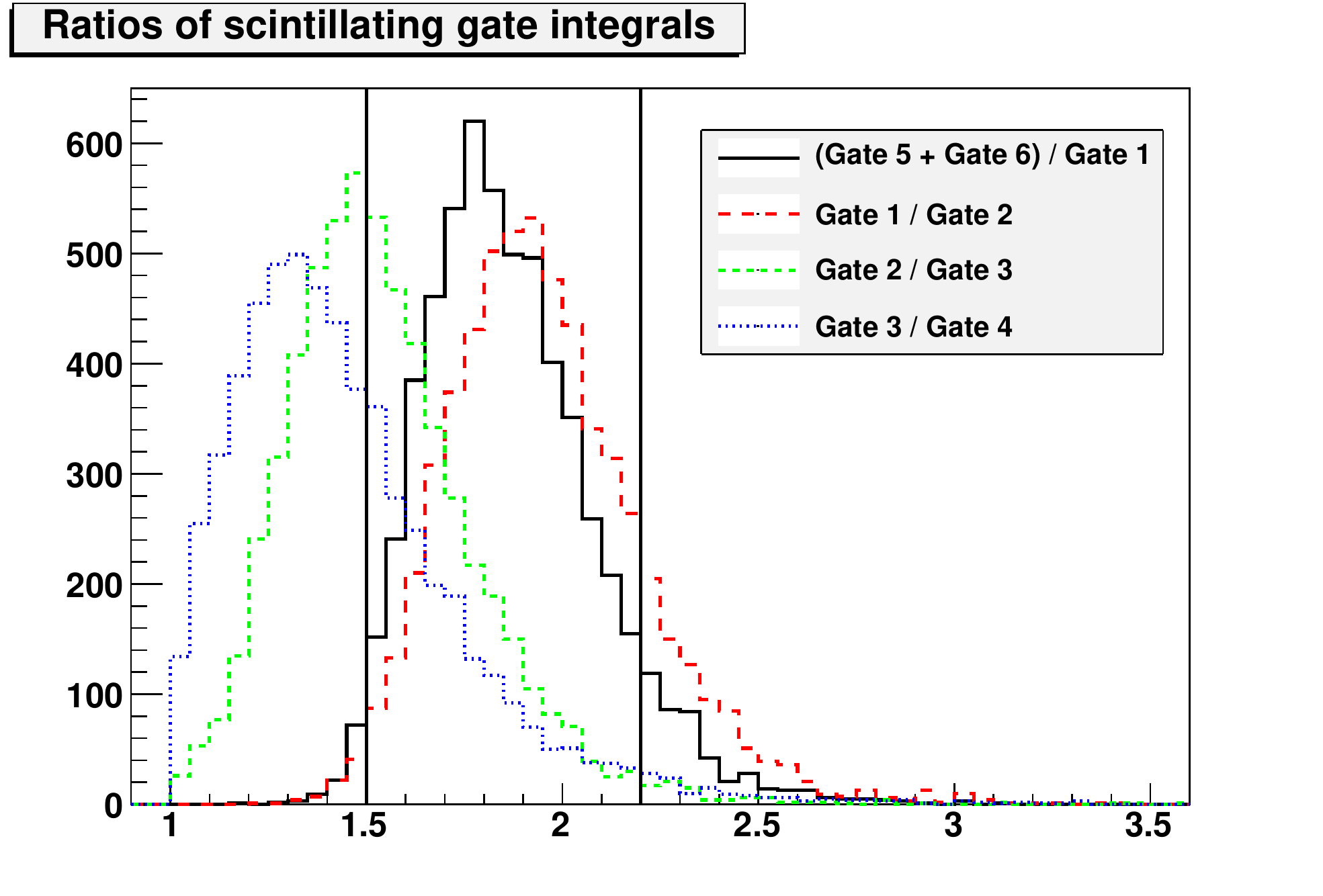}
\parbox{8.5cm}{\vspace{5pt}\caption{\small{Gate ratios for the scintillating-plastic detector. The curves' asymmetry prevented us from fitting Gaussian curves to the data. We incorporated manual ratio limits, however, that included at least 90\% of the curves' areas. For example, for the (Gate 5 + Gate 6) / Gate 1 ratio, we used limits of 1.5 and 2.2 (shown by the vertical bars), incorporating 92\% of the available area. Analogous limits were identified for the other three ratios.}}
\label{fig:NeutronScintGateRatios}}
\end{figure}

\begin{figure}[t!!!]
\centering
\subfigure{\includegraphics[width=8.5cm] {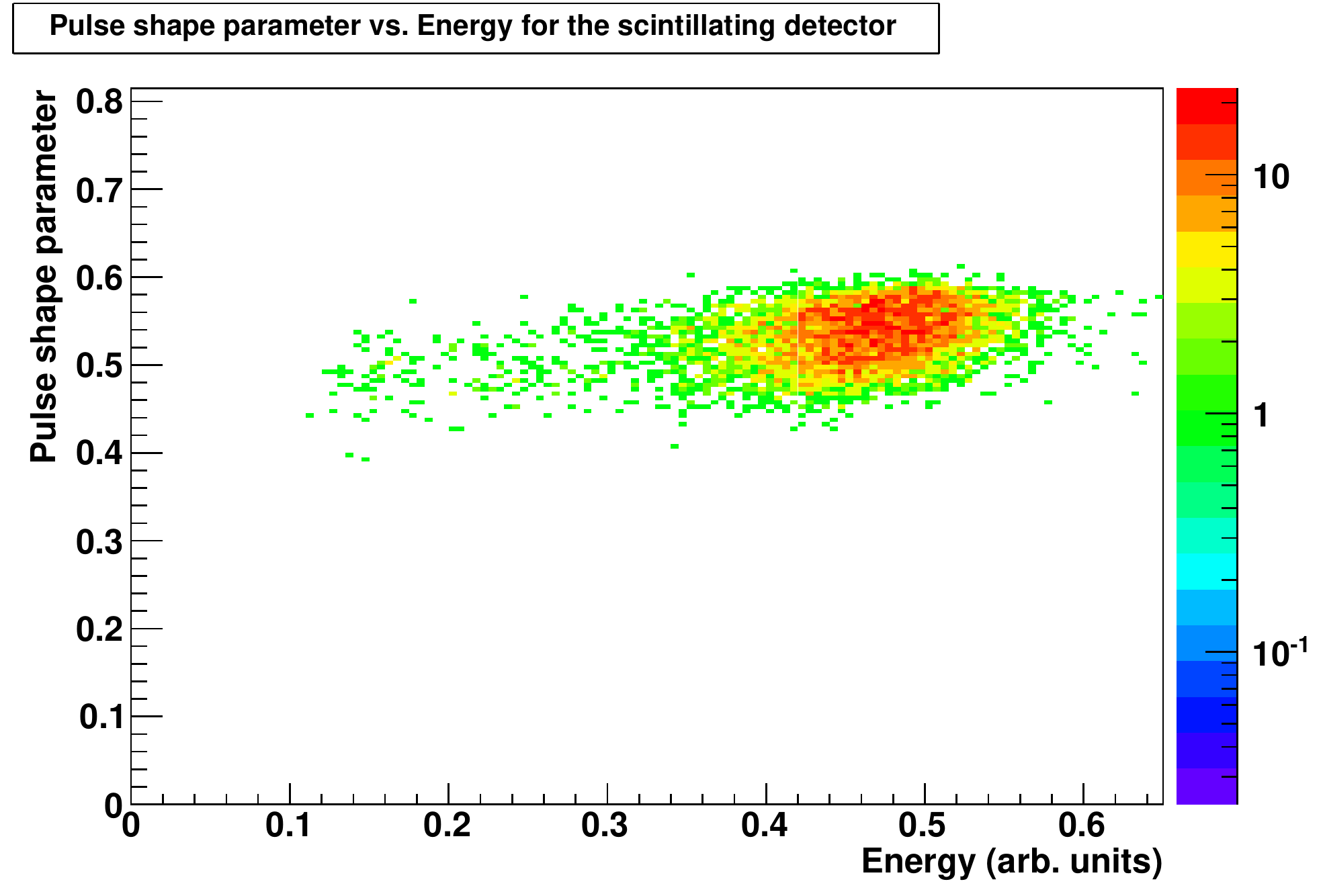}}
\subfigure{\includegraphics[width=8.5cm] {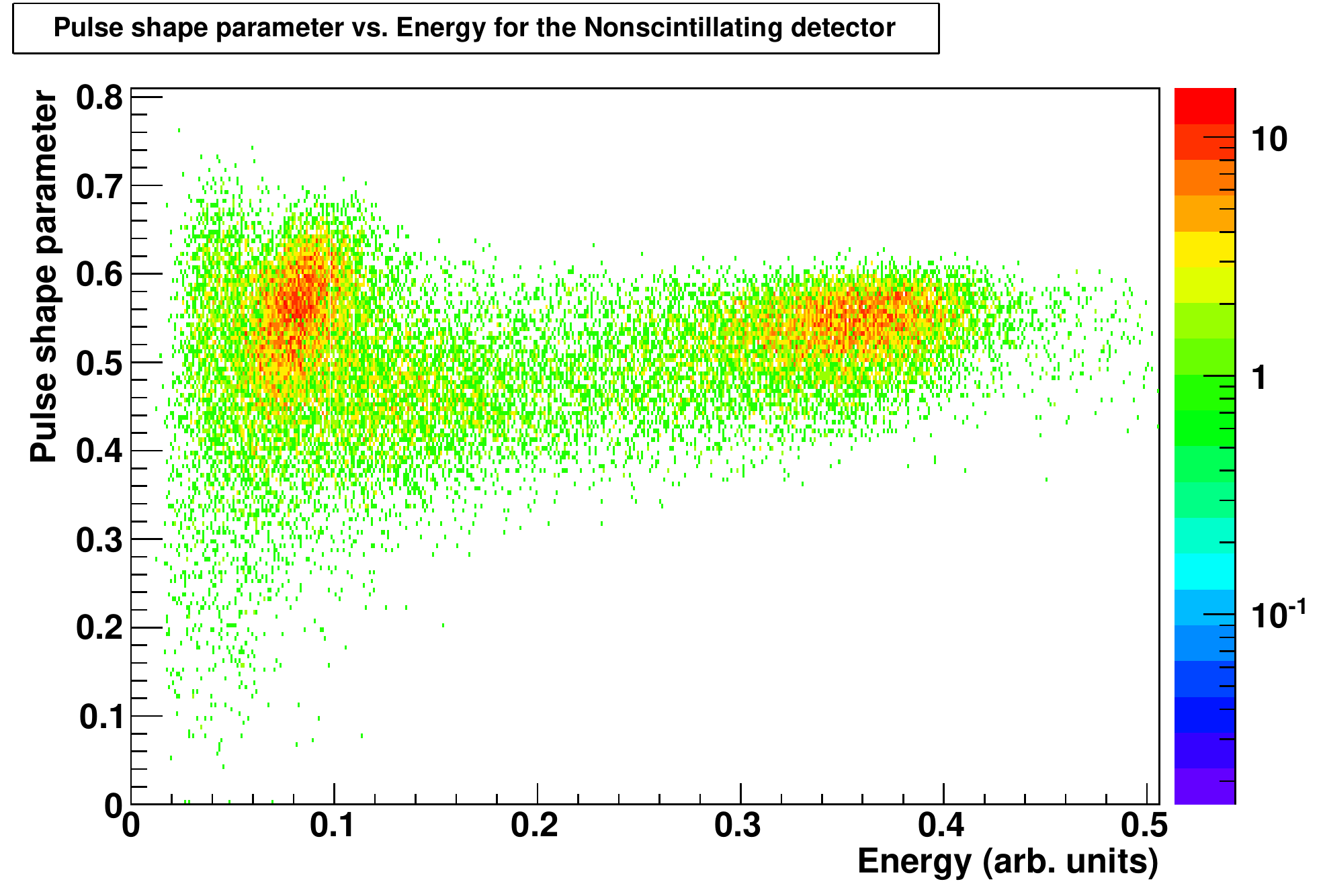}}
\parbox{8.5cm}{\vspace{5pt}\caption{\small{Data after applying the updated pulse shape discrimination procedure. Compare to Fig.~\ref{fig:NewAnalogResults}. The \iso{6}{Li} capture regions are unattenuated, while the gamma backgrounds have been drastically reduced. Note the second island in the lower plot, which is associated with captures on \iso{10}{B}. This island does not appear in the upper plot because of the 0.48~MeV gamma that accompanies \iso{10}{B} captures.}}
\label{fig:BetterPSA}}
\end{figure}

\noindent
We measured these ratios using events from the clearly-defined \iso{6}{Li}-capture islands from Fig.~\ref{fig:NewAnalogResults} for both the scintillating-plastic and nonscintillating-plastic detectors. Fig.~\ref{fig:NeutronScintGateRatios} shows the distribution of gate ratios as measured using the scintillating-plastic data. We then re-analyzed the full data sets, requiring all events have gate ratios as per the \iso{6}{Li}-capture islands. We then plotted the remaining events on a two-dimensional histogram analogous to those in Fig.~\ref{fig:NewAnalogResults}--with the simple pulse shape parameter on the vertical axis--for direct comparison between the simple and updated pulse shape discrimination procedures. These plots are shown in Fig.~\ref{fig:BetterPSA}. The success of our approach emphasizes the requirement that the LGB have a sufficiently long scintillation decay time.

Even a cursory comparison of Fig.~\ref{fig:NewAnalogResults} and Fig.~\ref{fig:BetterPSA} shows the efficacy with which the updated pulse shape discrimination removes background events.

\clearpage
\subsection{Measurements of neutron and gamma sensitivity}
\label{ss:Measurements}

We measured the intrinsic neutron capture efficiency and gamma sensitivity of our two detectors by acquiring three data sets. The first was the data set described in Section~\ref{ss:PSA}: a 20-hour data set with a bare \iso{252}{Cf} source approximately 3~feet from the detectors. The second data set was a 20-hour data set with a 14.0-kBq \iso{228}{Th} source, also suspended approximately 3~feet from the detectors. The final data set was a 20-hour background data set with no sources present. All three data sets had live times over 99\%. The maximum trigger rate was less than 10~kHz, ensuring that event pileup was not a concern. Given the source activities, length of data accumulation, and source distance from the detectors, we calculated a total neutron flux of $9.60 \times 10^5$ from the \iso{252}{Cf} source and a 2614-keV gamma flux of $1.39 \times 10^9$ from the \iso{228}{Th} source.

We would like to comment on the use of a \iso{228}{Th} to measure the effect of gamma rays on our composite detectors. Typically, a \iso{137}{Cs} source would be used to make such a measurement. Given the 2.2-MeV energy equivalent from a neutron capture on \iso{6}{Li}, the 662-keV gamma rays from \iso{137}{Cs} would virtually never create a background to the neutron capture signal, even if pulse shape discrimination were not used. Three gamma rays would have to interact within the LGB shards simultaneously, and deposit nearly 100\% of their energy within the shards. The probability of this is vanishingly small, even in a high-rate environment. To have a reasonable expectation of {\it any} gamma background at all, we needed to use a source with gamma rays at least 2.2~MeV. Of course, the gamma rays resulting from neutron captures on hydrogen are one possible source, but we wanted to separate the neutron and gamma signals. Thus we made use of the \iso{228}{Th} source, which contains within its decay chain \iso{208}{Tl}, and its 2614-keV gamma ray.

In the end, our gamma ray sensitivity measurements were more stringent than typical gamma sensitivity measurements performed on, for example, \iso{3}{He} proportional counter tubes.

\begin{figure}[t!!!]
\centering
\subfigure{\includegraphics[width=8.5cm] {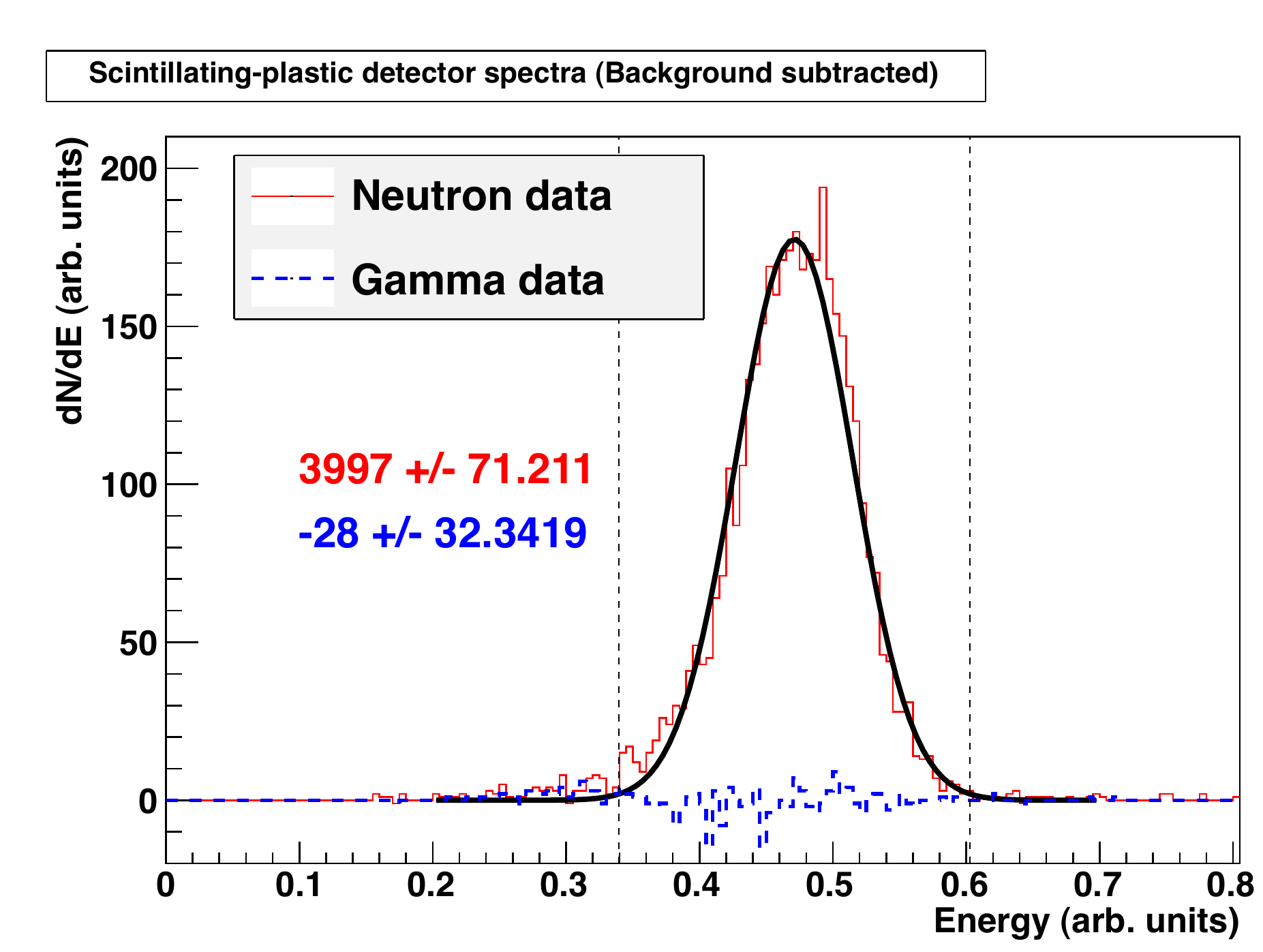}}
\subfigure{\includegraphics[width=8.5cm] {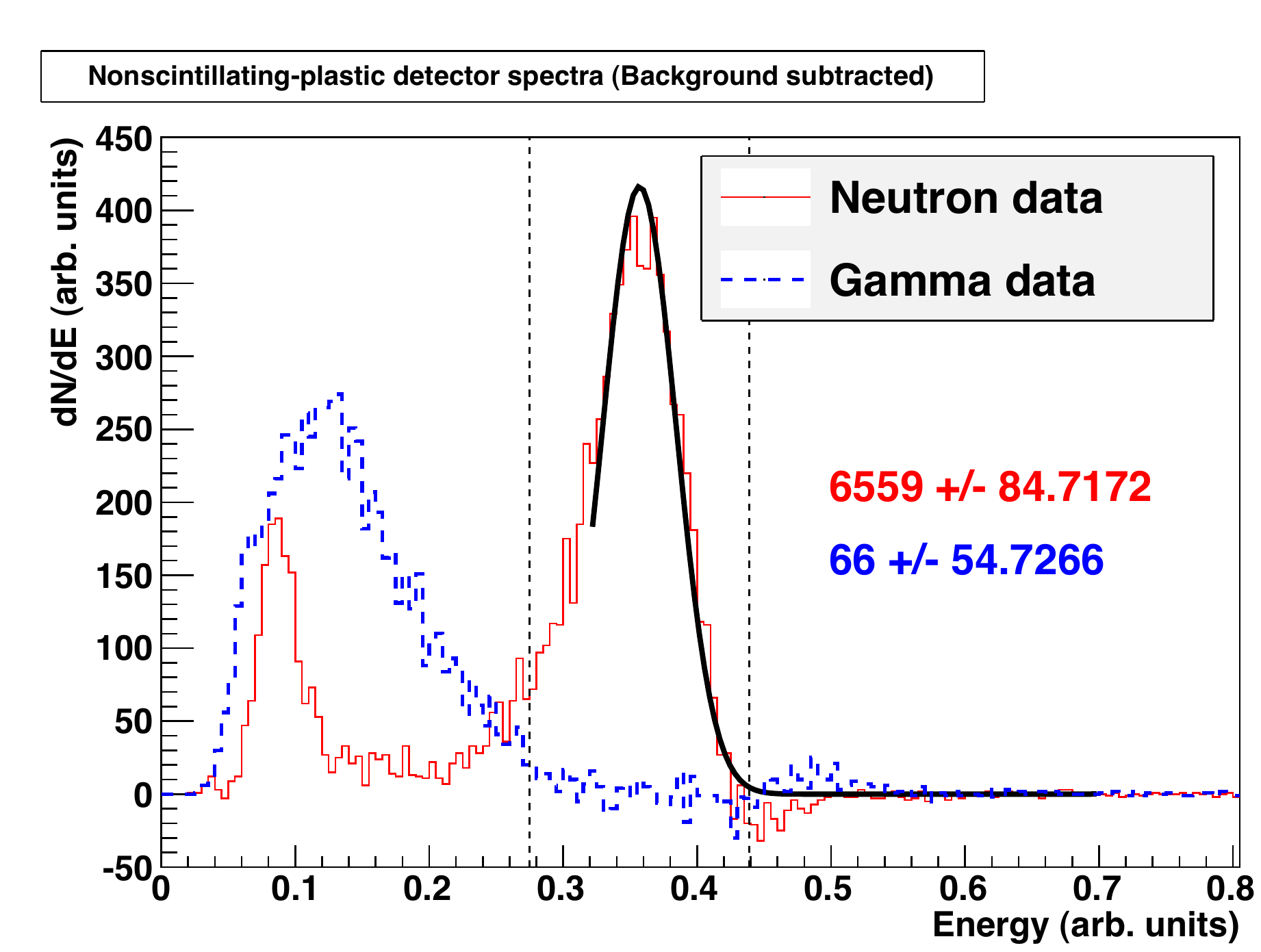}}
\parbox{8.5cm}{\vspace{5pt}\caption{\small{Background-subtracted data for neutron and gamma sources. We fit Gaussian curves to the neutron data, and integrated over $\pm3 \sigma$ for both the neutron and gamma curves. The integrals are shown on the plots. Because there was no normalization applied to the curves before subtracting, the integrals summed to integer counts. For both plots, the dashed vertical lines show the region of integration.}}
\label{fig:FinalMeasurements}}
\end{figure}

For all three data sets we applied the updated pulse shape discrimination, then utilized an additional cut in the simple tail / full ratio of between 0.4 and 0.6 for the scintillating-plastic detector, and between 0.4 and 0.7 for the nonscintillating-plastic detector. We projected the resulting data onto the x-axis and performed a bin-by-bin subtraction of the background histogram from both the gamma and neutron histograms. The results are shown in Fig.~\ref{fig:FinalMeasurements}.

Normalizing the integrals in the neutron signal regions by the total flux of either neutrons or 2614-keV gammas, we calculated the intrinsic neutron efficiency and gamma sensitivity of these detectors. Results are shown in Table~\ref{tab:FinalMeasurements}.

\begin{figure}[t!!!]
\centering
\includegraphics[width=8.5cm]{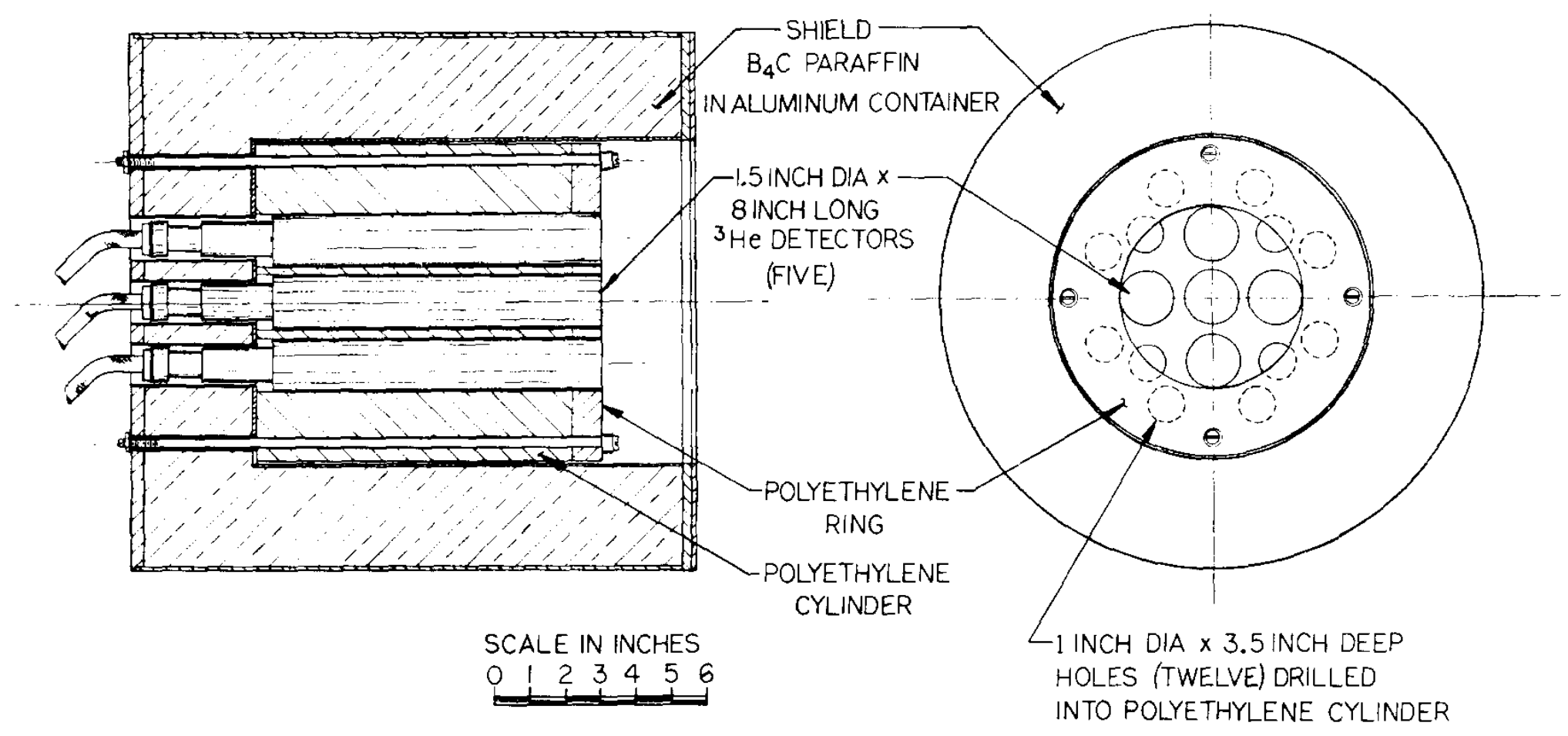}
\parbox{8.5cm}{\vspace{5pt}\caption{\small{Moderated \iso{3}{He} tube array for fast neutron detection. East and Walton measured an intrinsic fast neutron efficiency to be ($11.5 \pm 0.5$)\%. The central active volume is approximately 5" in diameter, the same as our composite LGB detectors, allowing for easy comparison between the detectors. Figure taken from Ref.~\cite{East1969}.}}
\label{fig:Moderated_He3_Setup}}
\end{figure}

We can compare the intrinsic efficiency of our LGB composite detectors to fast neutrons with that of moderated \iso{3}{He} tubes. East and Walton constructed an array of five \iso{3}{He} tubes~\cite{East1969}. Each tube was 8" long, 1.5" in diameter, and pressurized to 6 atmospheres. The tubes were embedded in a polyethylene moderator 9" in diameter, as shown in Fig.~\ref{fig:Moderated_He3_Setup}. The five tubes were fed to a single preamplifier, and the total intrinsic efficiency for fast neutrons was measured to be $(11.5 \pm 0.5)$\%. Our 5" detectors would fit very cleanly into the 9" cylinder, providing for 2" of polyethylene moderator. Menaa {\it et al.} measured a 10-fold increase in fast neutron efficiency for their LGB composite detector simply by using 2" of polyethylene moderator~\cite{Menaa2009}. We would therefore expect our nonscintillating-plastic detector to have an intrinsic efficiency for fast fission neutrons of $\sim$7\%. Increasing the performance of our detectors by a factor of two would make them competitive with \iso{3}{He}-based systems.

\begin{table}[t!!!]
\caption{\small{Intrinsic gamma and fission neutron sensitivity of the two detectors. Note that because the gamma curve in the scintillating-plastic detector integrates to less than 0, we can only apply a 1$\sigma$ upper limit on gamma sensitivity.}}
\begin{center}
\begin{tabular}{|c|c|c|}
\hline
Detector & Neutron sensitivity & Gamma sensitivity \\
\hline
Scintillating-plastic & (0.416 $\pm$ 0.007)\% & $< 2.3 \times 10^{-9}$ \\
Nonscintillating-plastic & (0.683 $\pm$ 0.009)\% & $(4.75 \pm 3.94) \times 10^{-9}$ \\
\hline
\end{tabular}
\end{center}
\label{tab:FinalMeasurements}
\end{table}

%
%
\clearpage
\section{Optimizing Crystal Size and Content}
\label{s:Optimizations}

The detectors utilized in this work were intended to demonstrate the feasibility of composite detectors, and as such they were not optimized for neutron sensitivity or gamma rejection. In this section we discuss possible optimizations to increase neutron sensitivity while maintaining or reducing gamma sensitivity.

\subsection{Crystal size}
\label{ss:CrystalSize}

\begin{figure}[b!!!]
\centering
\includegraphics[width=8.5cm]{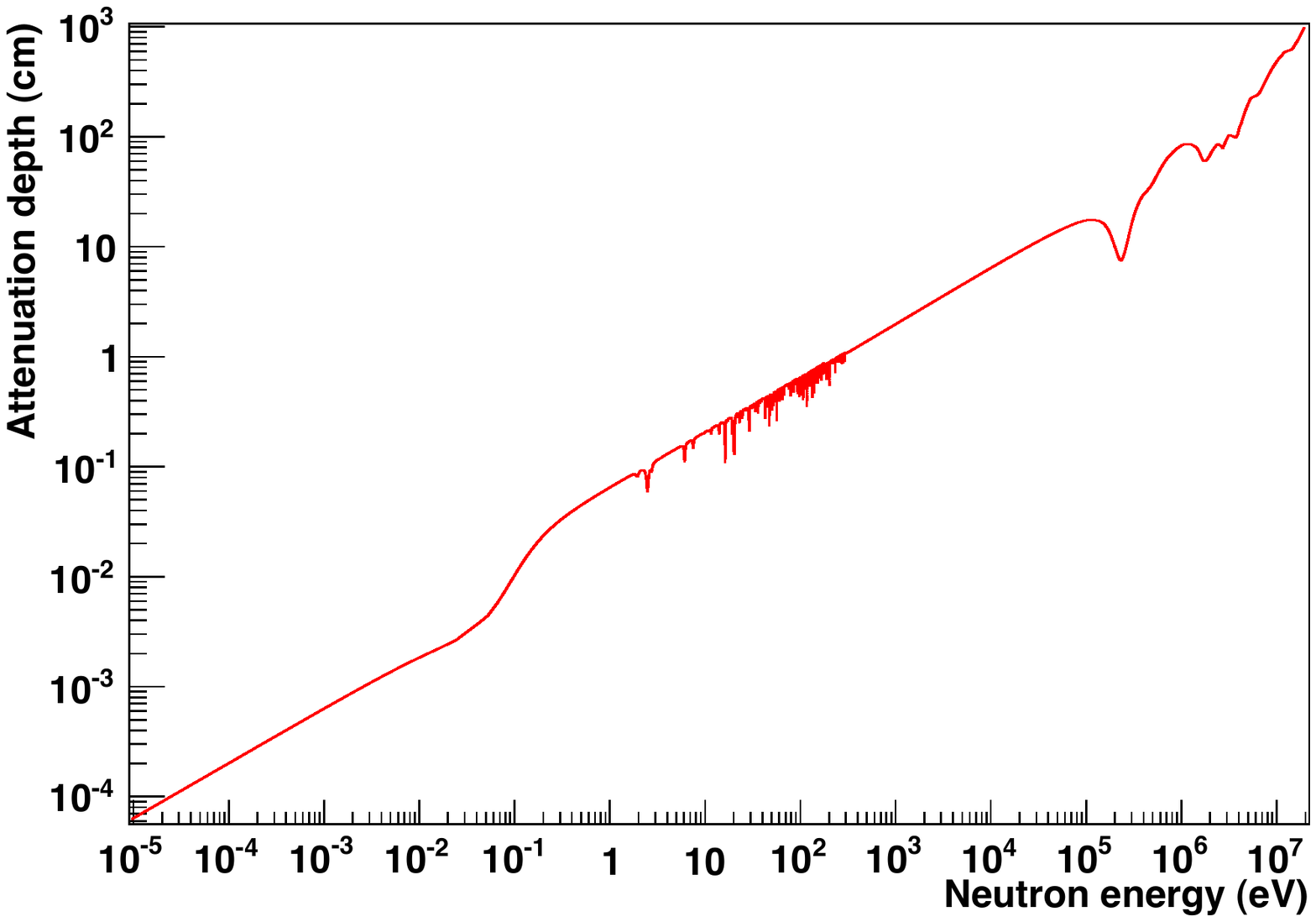}
\parbox{8.5cm}{\vspace{5pt}\caption{\small{Attenuation depth for neutrons in the LGB crystal. This crystal, as described above, is enriched to 95\% in both \iso{6}{Li} and \iso{10}{B}.}}
\label{fig:AttenuationDepth}}
\end{figure}

We begin the optimization analysis from the point of view of crystal self-shielding. The LGB crystals provide a signal when a neutron captures on the \iso{6}{Li} nuclei, and the capture depth of the neutrons is shown in Fig.~\ref{fig:AttenuationDepth}. At thermal energies, the capture depth is on the order of 10-20~$\mu$m, although neutrons are not necessarily completely thermalized by the time they capture. Regardless, the capture depth will tend to be small compared to the typical shard dimension of 1~mm, so the shards will exhibit considerable self-shielding. The captures occur primarily in the outer edge of the crystals, and the inner volumes are more or less unused.

From the point of view of self-shielding, the smaller the crystal shards, the more efficiently the LGB crystals capture neutrons, even with the same total mass. Recall, however, that the recoiling ions have a range of approximately 30~$\mu$m, so if the crystals become too small, we again potentially lose efficiency because less scintillation light will be produced in the shard. There are therefore multiple parameters that need to be considered when calculating the optimal shard dimension, including the primary neutron spectrum, efficiency of thermalization, capture cross-section of both the plastic and the LGB crystal, angle of ion recoil with respect to the shard surface, and so on. These multiple related variables make an analytic calculation intractable, so we turn to a Monte Carlo calculation to determine the optimal shard dimension.

\begin{figure}[b!!!]
\centering
\includegraphics[width=8.5cm]{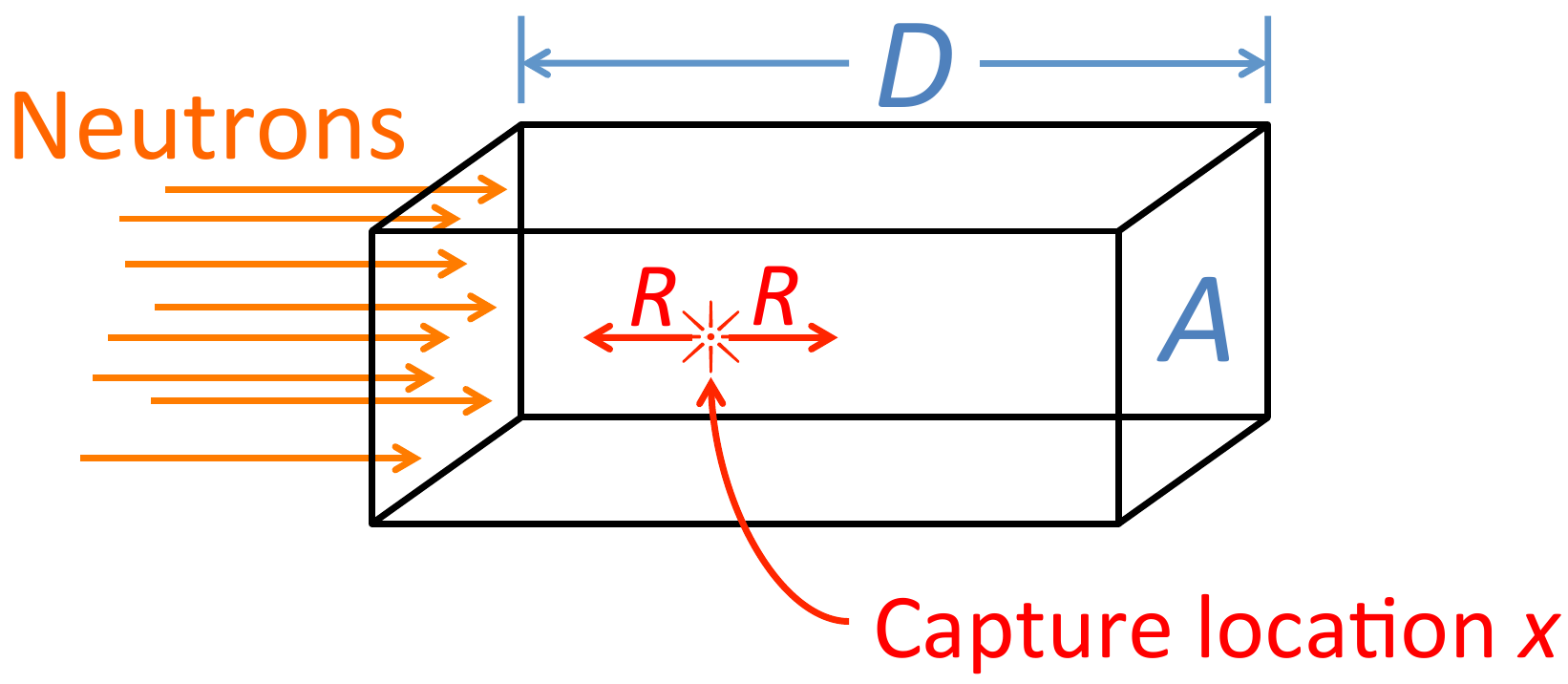}
\parbox{8.5cm}{\vspace{5pt}\caption{\small{Setup for calculating a simple, one-dimensional optimization of the LGB shard size. $D$ is the length of the crystal, and $A$ is its cross-sectional area. Thus $D$ and $A$ are inversely proportional. $R$ is the range of the recoil ions. Because this is a one-dimensional calculation, the angle of ion recoil is irrelevant.}}
\label{fig:SimpleOptimization}}
\end{figure}

To determine an appropriate range over which to run the Monte Carlo calculations should be run, we perform a simple one-dimensional calculation to predict the optimal dimension. We assumed a constant-mass LGB crystal with a uniform flux of monoenergetic neutrons normal to one face of the crystal (Fig.~\ref{fig:SimpleOptimization}). $D$ is the depth of the crystals along the direction of the incident neutron flux.

We define $\lambda$ as the capture depth of a neutron, which in this treatment is taken to be a constant, in accordance with the simplification of monoenergetic neutrons. The capture probability for a single neutron entering the crystal is therefore $1 - e^{-D/\lambda}$. The total number of neutrons entering the crystal is the neutron flux multiplied by the cross-sectional area $A$. The smaller the value for $D$, the larger the total flux. The total capture efficiency therefore goes as

\begin{equation}
\mbox{Efficiency} \sim \left ( 1 - e^{D/\lambda} \right ) / D
\label{eq:SimpleOptimization}
\end{equation}

\begin{figure}[b!!!]
\centering
\includegraphics[width=8.5cm]{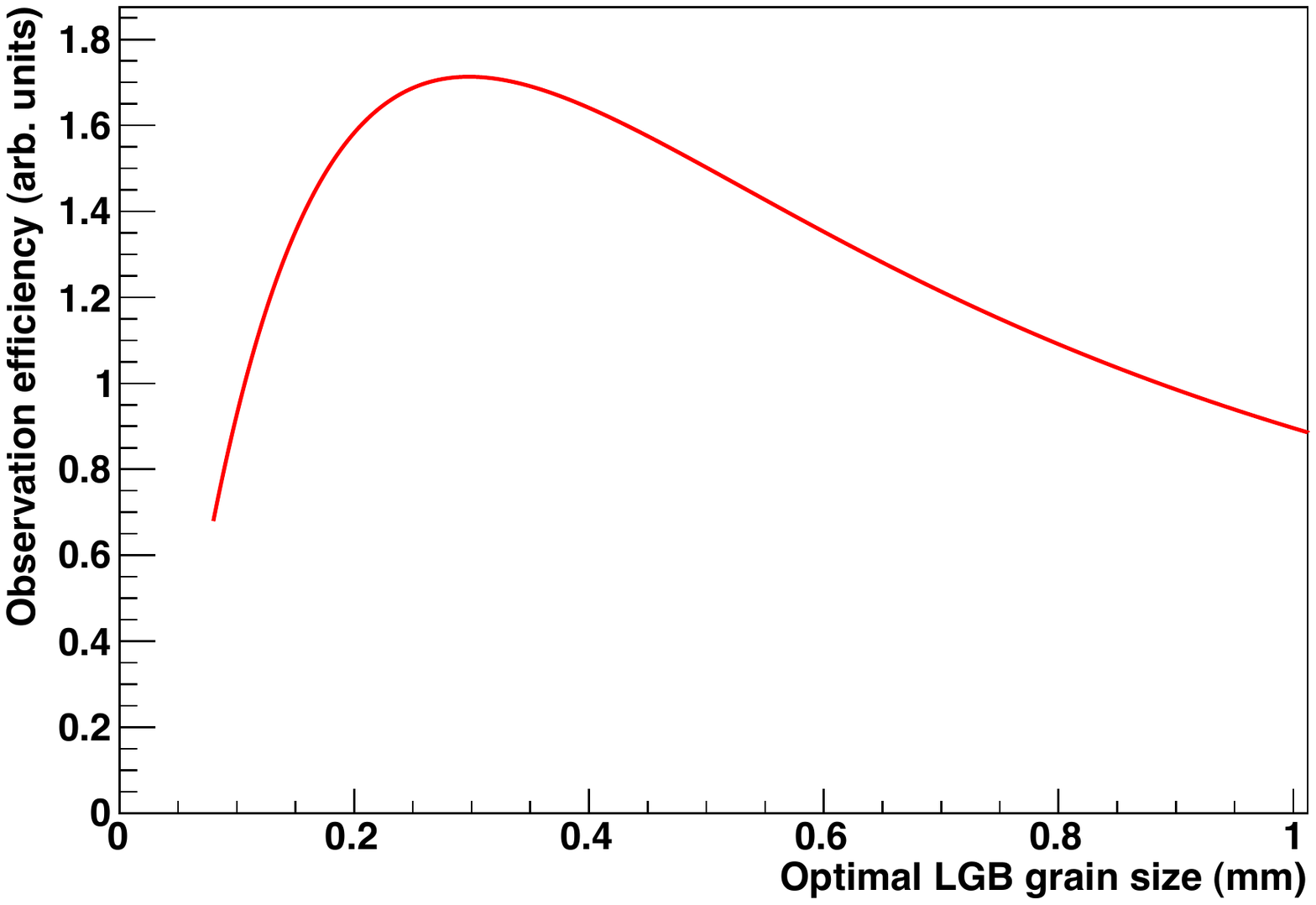}
\parbox{8.5cm}{\vspace{5pt}\caption{\small{Simple, one-dimensional optimization of the crystal shard size for $R = 40~\mu$m and $\lambda = 200~\mu$m. The maximum occurs near 300 $\mu$m, so our Monte Carlo calculations will approximately span the range 100~$\mu$m to 1~mm.}}
\label{fig:SimpleOptimizationResults}}
\end{figure}

Next, we note that if the capture location within the crystal is greater than a distance $R$ from either end of the crystal, the recoiling ions will deposit 100\% of their energy in the crystal. Similarly, if $D < 2R$, the ions will never deposit full energy in the crystals. We therefore integrate Eq.~\eqref{eq:SimpleOptimization} over the capture location, $x$, from 0 to $D$ to obtain the probability of  the recoiling ions depositing their full energy in the crystals, assuming a minimum value of $2R$ for $D$. Assuming conservative values of 40~$\mu$m for $R$ and 200~$\mu$m for $\lambda$, the efficiency curve is shown in Fig.~\ref{fig:SimpleOptimizationResults}. Within the limits of this simple approximation, we expect the optimal shard size to be on the order of hundreds of microns.

\subsection{Total crystal content}
\label{ss:TotalCrystalContent}

A second optimization to be performed using Monte Carlo calculations is the total crystal content of the composite detector. The detectors used in this study were 1\% by mass LGB, and increasing crystal content would increase neutron sensitivity. Increasing crystal content, however, might also increase gamma sensitivity, and thus the Monte Carlo calculations must be run with both fission neutrons and gamma rays.

The effects of total crystal content on both neutron and gamma sensitivity can be somewhat subtle. If the crystal content is low, the neutron sensitivity should be expected to scale linearly with it. The gamma sensitivity, however, is a strong function not only of total crystal content, but crystal dimension as well. Recall from Section~\ref{s:Intro} that the range of a 2.2-MeV electron in the LGB crystal is 4.4~mm, while the range of the tritium ion from capture on \iso{6}{Li} is only about 30~$\mu$m. This implies that if the shards are small, the gamma background to the neutron capture signal can be kept small. Unfortunately, if the crystal content is sufficiently large, it is possible for a recoiling electron to traverse multiple LGB shards. In this case, even if each shard emits only a fraction of the light equivalent of a 2.2 MeV of electron energy, the sum total of the light emitted by all the shards may create a background to the neutron signal.

\subsection{Monte Carlo Results}
\label{ss:MonteCarloResults}

For our Monte Carlo calculations, we used LUXSim, a simulations package based on Geant4~\cite{LUXSim}. Geant4 is a C++-based framework for simulating the effects of various kinds of radiation in different materials, driven by both physics models as well as data~\cite{Agostinelli2003}. Geant4 is capable of handling electrons, gammas, neutrons, and ions.

For the calculations, we used Geant4, version 4.9.3.p01. Unfortunately, there was also a documented bug in Geant4 that results in incorrect ion recoil energies resulting from neutron captures on \iso{6}{Li}~\cite{Hartwig2010}. For our simulations, we implemented a correction so that the recoiling tritium and helium nuclei had the correct energies. As a side note, there was also a bug for captures on \iso{10}{B}, although since our analysis only relied on captures on \iso{6}{Li}, the effects of the \iso{10}{B} bug could be safely ignored.

\begin{figure}[b!!!]
\centering
\includegraphics[width=6.5cm]{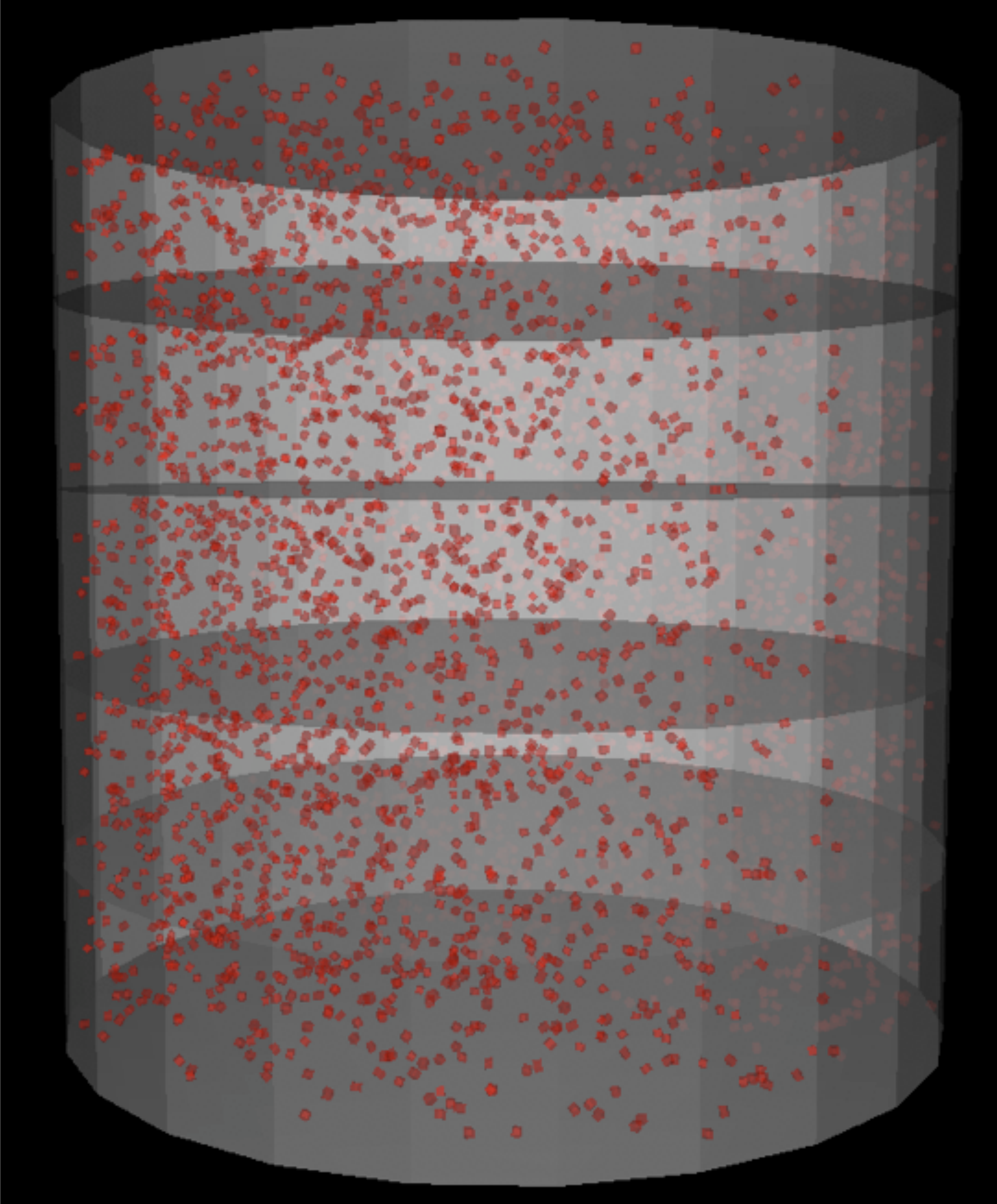}
\parbox{8.5cm}{\vspace{5pt}\caption{\small{Example Geant4 geometry of the composite detector. In this example, the shard size is 0.5~mm, the total content is 1\%, and there are 5 layers. Note that the shards, while having identical dimensions, are placed in the cylinder in random orientation. The layers themselves, though, have identical placement of the shards.}}
\label{fig:Geant4Cylinder}}
\end{figure}

The LGB crystal shards in our simulations were all placed individually inside the plastic cylinder. This was to ensure there was no overlap between the crystals, thereby guaranteeing the total crystal content is well-defined. Unfortunately, this resulted in an exceedingly large number of shards to place, each one having to be checked against overlap with any others. To make the placement more tractable, the cylinder was sub-divided into layers, and a proportionately smaller number of crystal shards were placed in the layer. The layer itself was then duplicated to create the full cylinder. An image of such a construction is shown in Fig.~\ref{fig:Geant4Cylinder}.

\begin{figure}[t!!!]
\centering
\includegraphics[width=8.5cm]{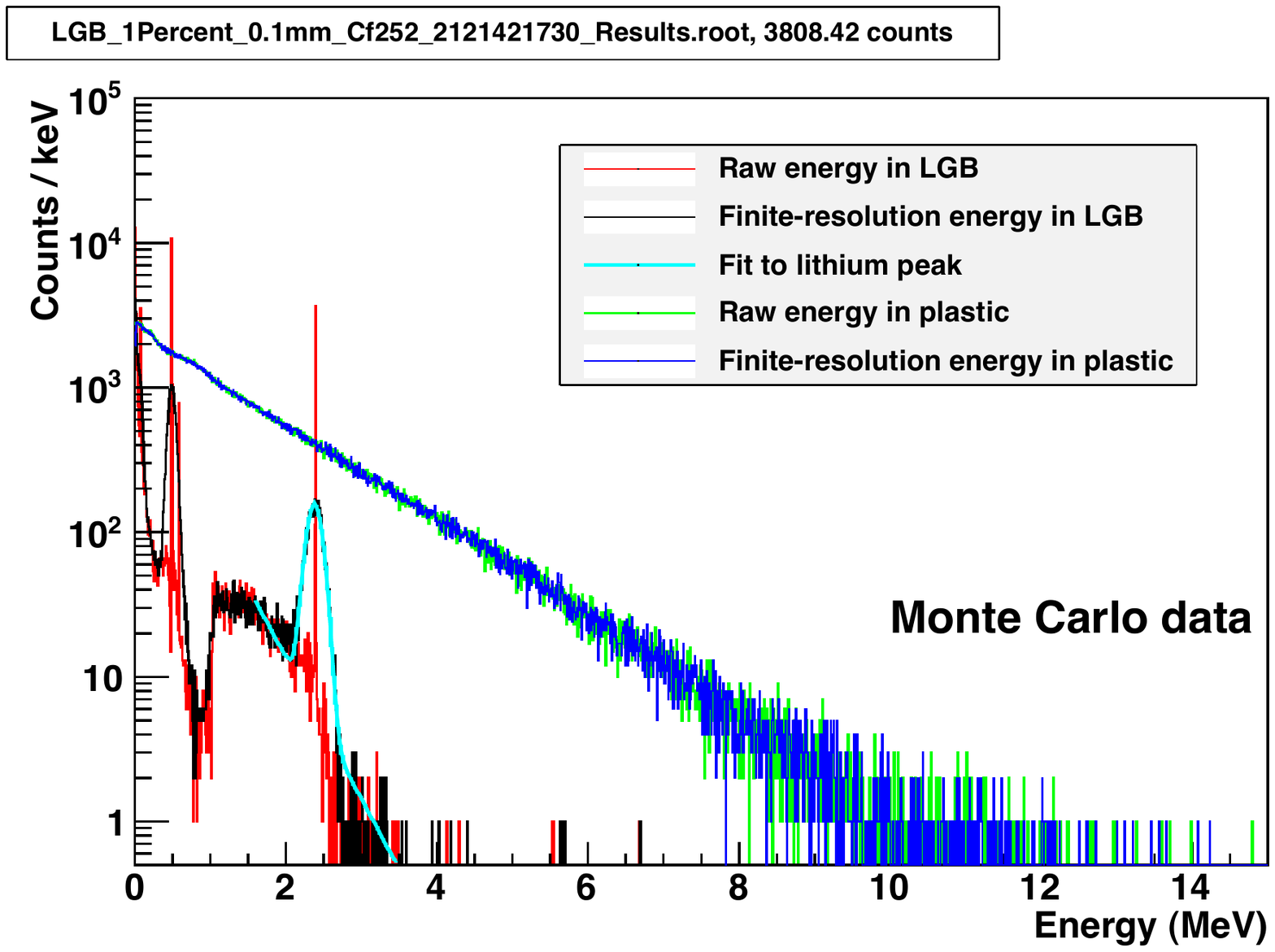}
\parbox{8.5cm}{\vspace{5pt}\caption{\small{Sample plot of analysis of a Geant4 simulation with a neutron source. The unquenched energy depositions range up to 10-12~MeV because of the sum of the energy of the primary neutron and the energy of the 8-MeV gamma cascade resulting from neutrons capturing on gadolinium nuclei. See text for details on each curve.}}
\label{fig:SampleAnalysisPlot}}
\end{figure}

For each simulation data set, we recorded the total energy deposition in the plastic and the crystal, as well as the particle that created that energy deposition. If the particles depositing the energy came from neutron captures on \iso{6}{Li}, their energy deposition was quenched by a factor of 0.46 (the ratio of 2.2~MeV / 4.78~MeV discussed in Section~\ref{s:Intro}). A finite resolution of 9\% was applied to the energy depositions, to match the width of the Gaussian peak in the upper plot of Fig.~\ref{fig:FinalMeasurements}. Finally, a 3$\sigma$ range around the centroid of the \iso{6}{Li} peak was defined, and any counts within this range was tabulated. An example of this entire process is shown in Fig.~\ref{fig:SampleAnalysisPlot}, which shows the results from a simulation with 1-mm crystal shards, and 1\% LGB content, with a fission neutron source.

\begin{figure}[t!!!]
\centering
\includegraphics[width=8.5cm]{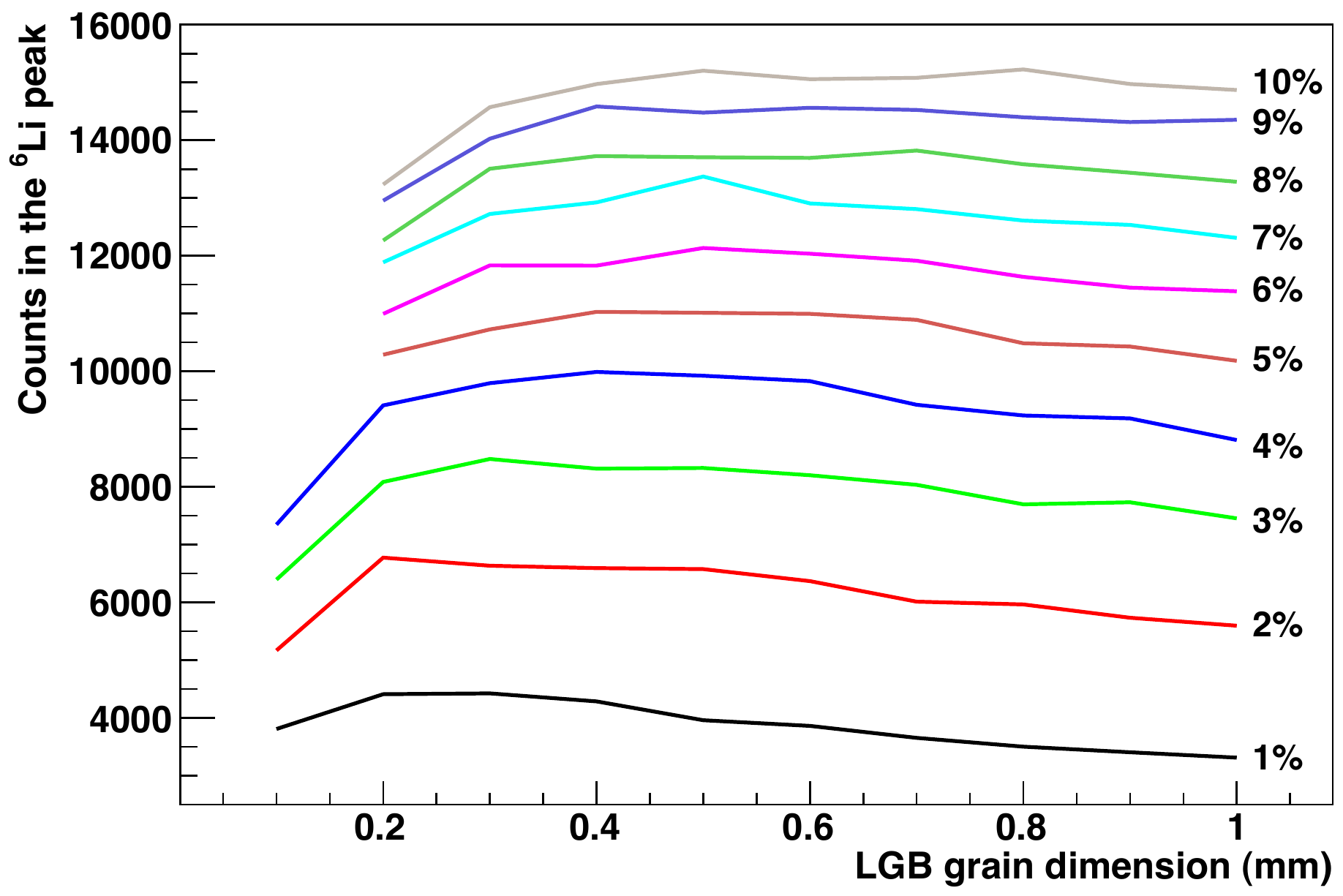}
\parbox{8.5cm}{\vspace{5pt}\caption{\small{Optimizing the shard size and total LGB content for the composite detector. Each curve represents the amount of LGB in the detector by mass. The detectors used in this study had a shard size of $\sim$1~mm, and a content of 1\%. Computing limitations prevented us from running simulations with 100-$\mu$m shard size above 4\% content.}}
\label{fig:LGBShardSizeOptimization_Neutrons}}
\end{figure}

We ran many simulations to fill out the parameter space of interest. We varied the primary particles between \iso{252}{Cf} fission neutrons, and \iso{208}{Tl} decays to provide the 2614-keV gamma rays from the \iso{228}{Th} decay chain. We varied the shard dimension from 100~$\mu$m to 1~mm in 100-$\mu$m steps, and we varied the total mass of the LGB crystals from 1\% to 10\% by mass, in 1\% steps. The results for the fission neutron source are shown in Fig.~\ref{fig:LGBShardSizeOptimization_Neutrons}. In this figure, we see that while the curves are much flatter than the simple curve from Fig.~\ref{fig:SimpleOptimizationResults}, the optimal shard size lies generally in the 200-$\mu$m to 600-$\mu$m range. We also note that, based on these Monte Carlo simulations, we can increase the neutron sensitivity by roughly a factor of 5 by using 10\% LGB content and 500-$\mu$m-sized shards. We note also, though, that given the somewhat flat response, the intrinsic neutron efficiency is somewhat insensitive to precise shard dimensions, provided the shards themselves are larger than $\sim$300~$\mu$m.

\begin{figure}[t!!!]
\centering
\subfigure{\includegraphics[width=8.5cm] {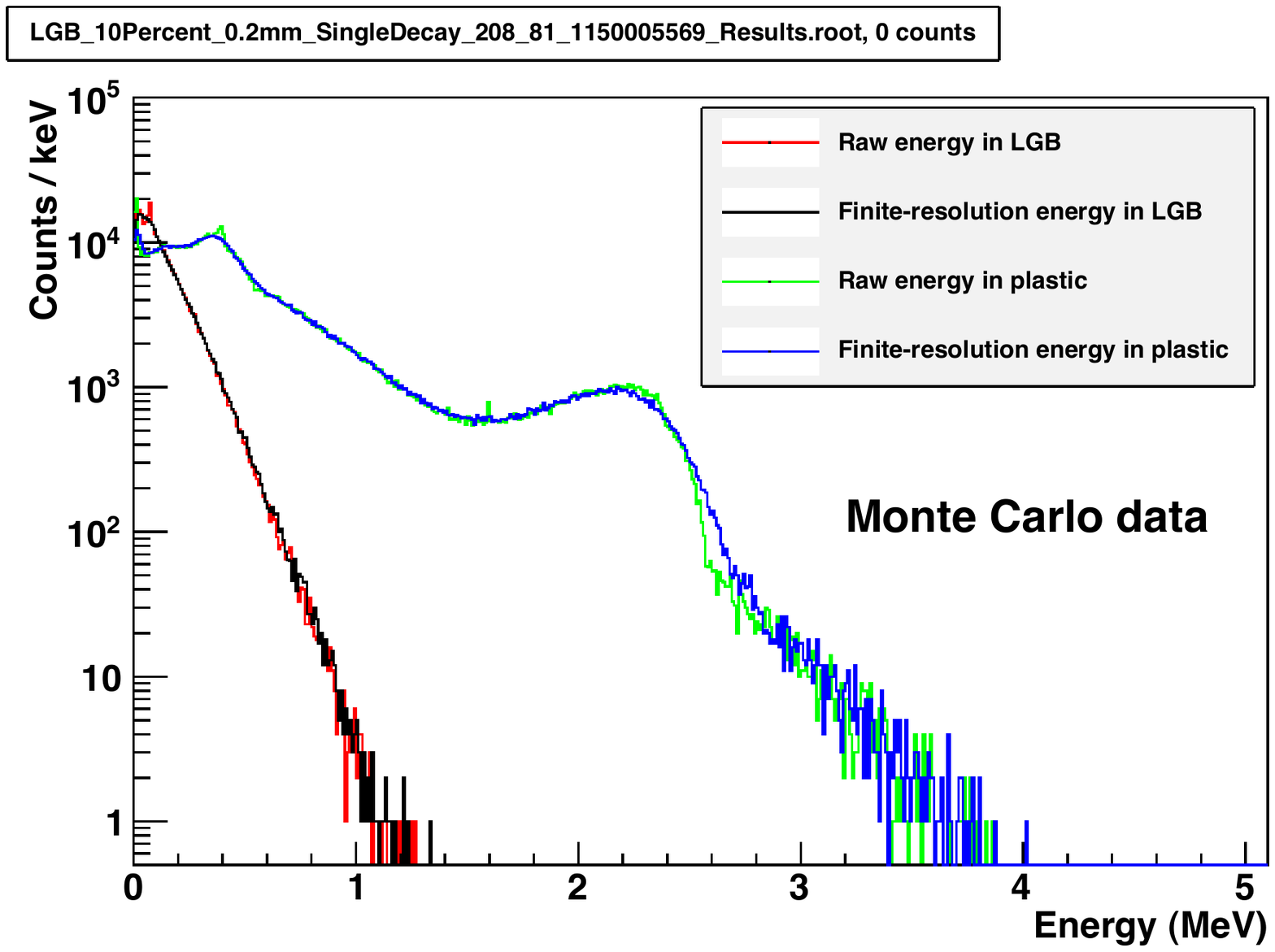}}
\subfigure{\includegraphics[width=8.5cm] {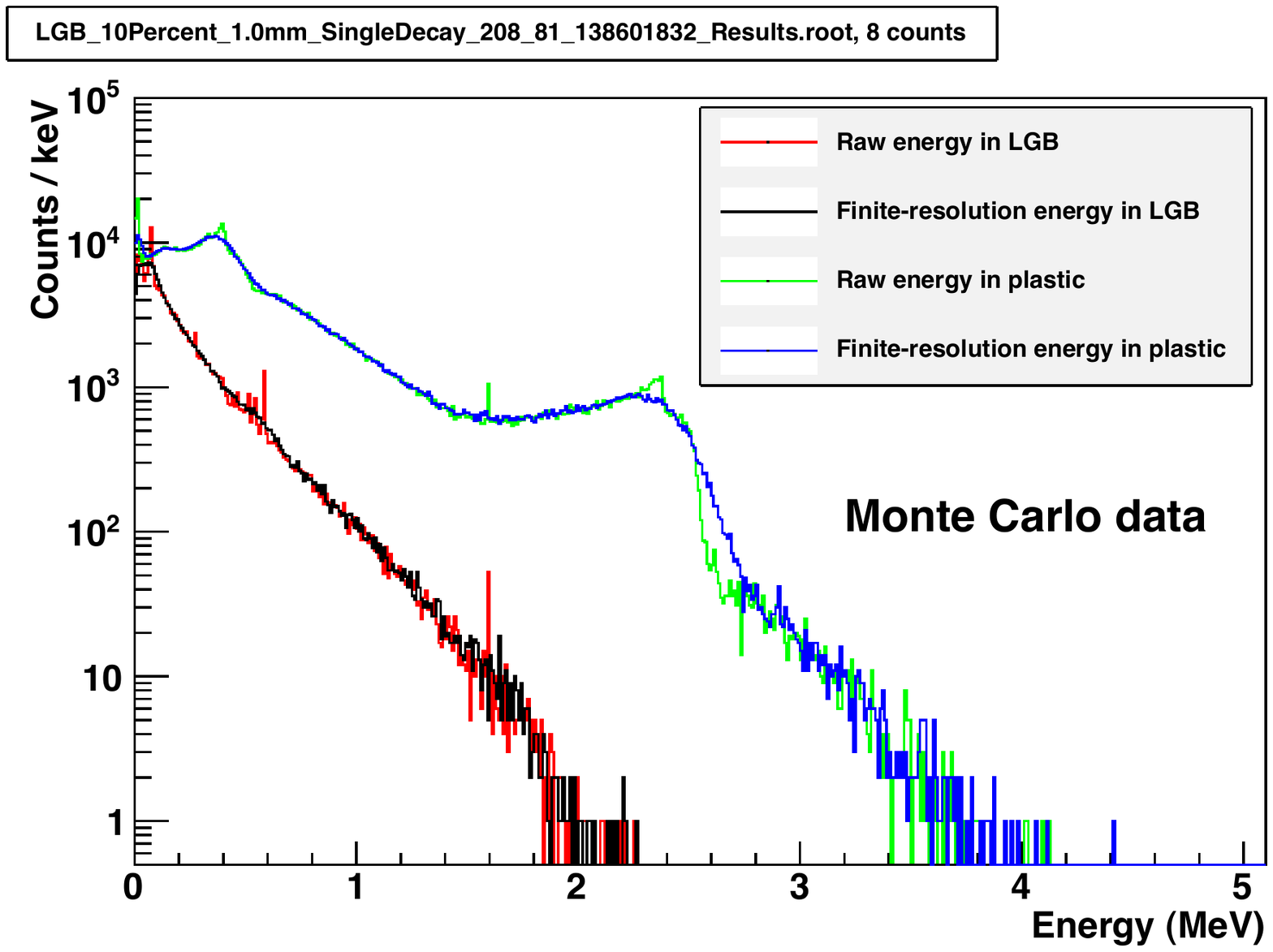}}
\parbox{8.5cm}{\vspace{5pt}\caption{\small{Sample analysis plots of the Geant4 simulations with the \iso{208}{Tl} source. For both of these plots, the total LGB content was 10\%. For the top plot, however, the shard size was 0.2~mm, and for the bottom plot the shard size was 1~mm. We see that the shard size can greatly affect the number of gamma counts in the \iso{6}{Li} neutron capture region. Indeed, the top plot had zero counts in this region, while the bottom plot had 8 counts. The maximum possible total energy in the gamma rays from \iso{208}{Tl} decays is 4.4~MeV, although the branching ratios to those levels are very small. The branching ratio to the 3.96-MeV energy level is 3\%, which is the approximate endpoint of the energy depositions in the plastic.}}
\label{fig:SampleGammaAnalysisPlot}}
\end{figure}

\begin{figure}[t!!!]
\centering
\includegraphics[width=8.5cm]{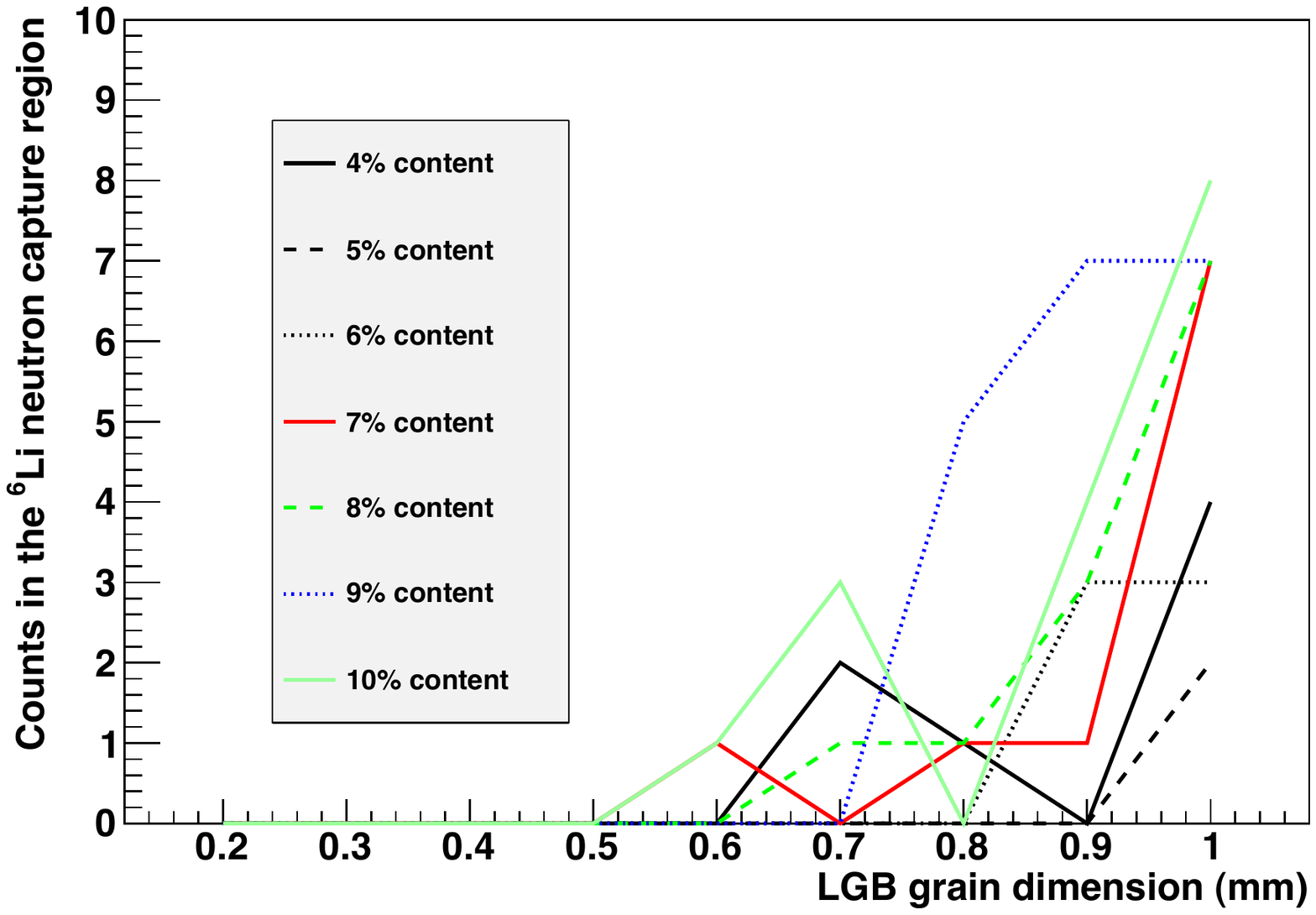}
\parbox{8.5cm}{\vspace{5pt}\caption{\small{Gamma sensitivity in the \iso{6}{Li} neutron capture region. This plot is analogous to Fig.~\ref{fig:LGBShardSizeOptimization_Neutrons}, but with a \iso{208}{Tl} source instead of a fission neutron source. There were no counts at all for LGB content lower than 4\%. While the statistics are low, it is clear to see that, regardless of the LGB content, reducing the shard size from 1~mm to 0.5~mm will not increase the gamma sensitivity.}}
\label{fig:GammaBackground}}
\end{figure}

The question remains, however, of the effect the crystal size and content on the gamma sensitivity. Using the \iso{208}{Tl} data, we counted the number of events in the \iso{6}{Li} neutron capture region. Sample analysis plots similar to that in Fig.~\ref{fig:SampleAnalysisPlot} are shown in Fig.~\ref{fig:SampleGammaAnalysisPlot}. They show that, even with the same total LGB content, the shard size can still have a substantial effect on the potential gamma sensitivity to these composite detectors. The final plot from our gamma sensitivity simulations is shown in Fig.~\ref{fig:GammaBackground}. From this plot we conclude that at the optimal shard size for neutron captures, 500~$\mu$m, the gamma sensitivity is consistently reduced relative to a shard size of 1~mm. This gives us confidence that the LGB content can be increased to the 10\% range without increasing gamma sensitivity.

One final note on crystal optimization is the effects of small shards on optical transparency. The LGB has an index of refraction of 1.66, compared to the EJ-290's value of 1.58. Because of this refractive index mismatch, a plastic matrix with a high LGB content may be optically cloudy (see, e.g., Fig. 1 from Ref.~\cite{Menaa2009}, where the detectors were 27\% LGB by mass). A full treatment of the optimization of LGB content would have to include scintillation photon attenuation. This attenuation must be measured for any given final configuration, and extrapolations or interpolations made from there to determine optimal LGB content given the overall dimensions of the plastic matrix.

%
%
\clearpage
\section{Additional Improvements}
\label{s:Improvements}

In addition to optimizing the crystal content and shard size, we have identified other avenues of increasing detector performance. The goal is to increase neutron sensitivity while maintaining the very high level of gamma rejection.

\subsection{Changes to the data acquisition system}
\label{ss:DAQchanges}

One method is to change the gate sizes of the data acquisition software. This could be done through a firmware update, although the code itself would have to be contracted from Struck. Using the eight available gates, it would be beneficial to change their maximum sample sizes. One gate would be used to measure the baseline value on an event-by-event basis, and 50 samples would suffice. The LGB scintillation pulses only need to be digitized for 5 times the decay constant, or $\sim$1350~ns. Subdividing this pulse by 7 gives roughly 39 samples per gate. As it is, having 4 gates limited to 16 samples while 4 gates have 512 samples available means there is a fair amount of flexibility in shuffling around the available space. Certainly, 8 gates with a maximum sample size of 50 would suffice for this application.

With the 7 data points, one gate would cover the rising edge of the pulse. The other six samples could then be used to identify a very clean exponential decay. Indeed, to perform better pulse shape analysis, an exponential curve could be fit to the six decay data points, with the time constant a fixed parameter. The only free parameter in the fit would therefore be the amplitude of the pulse. The pulse shape discrimination would them come about via the reduced $\chi^2$ of the fit, where an associated p-value between, for example, 0.05 and 0.95 could be counted as a good fit. Anything else would be rejected as pileup, a gamma event, a muon event, or some other spurious background.

An additional benefit of this fitting approach would be the reconstruction of the full pulse integral. Once the fit amplitude is known, the energy resolution could conceivably increase, leading to a smaller neutron capture region, and therefore reduced background.

\subsection{Alternatives to lithium gadolinium borate}
\label{ss:LGBAlternatives}

The analysis in this work relied entirely on neutron captures on \iso{6}{Li} within the LGB crystal. Recall that the lithium was enriched in \iso{6}{Li} and the boron in \iso{10}{B}, while the gadolinium was of natural isotopic content. In Ref.~\cite{Nelson2011}, the authors showed that using gadolinium and boron depleted in their high-capture-cross-section isotopes, the number of captures on \iso{6}{Li} could increase by a factor of 6. While depleted boron is not difficult to obtain, depleted gadolinium, unfortunately, is.

We therefore identified materials other than LGB that could potentially be of use. These materials should have the following characteristics to make them suitable alternatives to LGB:

\begin{itemize*}\vspace{-0.5\baselineskip}
\item High lithium content
\item Long scintillation decay constant
\item No isotopes to compete with the lithium for neutron captures
\item High light output
\end{itemize*}

We have identified a few possible materials for future testing, shown in Table~\ref{tab:LGBAlternatives}. Of these materials, the most directly compatible with the results from this work is LYB, since it merely replaces the gadolinium with yttrium. Given this simple extrapolation, we estimate the lithium capture rate would increase by a factor of 6 over LGB, in accordance with results from~\cite{Nelson2011}. While the LYB light yield is small, this is not anticipated to be a strong issue, as the light yield on the scintillating plastic could be reduced a similar amount, while the gain on the PMTs could be turned up to compensate for the loss of light. If the refractive index of the plastic were matched to that of the inorganic crystal, the light collection would be further improved.

\begin{table}[t!!!]
\caption{\small{Possible alternatives to LGB.}}
\begin{center}
\begin{tabular}{|c|c|c|c|c|}
\hline
Material name & Composition & $\begin{array}{c} \mbox{Light Output} \\ \mbox{rel. to LGB} \end{array}$ & $\begin{array}{c} \mbox{Scint.} \\ \tau \mbox{~(ns)} \end{array}$ & Reference\\
\hline
LYB & Li$_6$Y(BO$_3$)$_3$ & 0.09 & 100 & \cite{Knitel2000} \\
Li glass & Variable & 0.29 & 62 & \cite{SaintGobain2007} \\
LiF & LiF(W) & 0.14 & 40000 & \cite{Pritychenko1997} \\
CLYB & Cs$_2$LiYBr$_6$ & 2.21 & 85 & \cite{Bessiere2004} \\
LTO & Li$_3$TaO$_4$ & 2.14 & 6100 & \cite{Derenzo2010} \\
LTB & LiB$_3$O$_5$ & 0.05 & $< 10$ & \cite{Nazarenko2006} \\
\hline
\end{tabular}
\end{center}
\label{tab:LGBAlternatives}
\end{table}

Lithium glass presents another attractive alternative, for two reasons. One is that its index of refraction ($n = 1.56$), is even closer to that of the plastic ($n = 1.58$) than LGB ($n = 1.66$). This would allow for greater light propagation, counteracting the lower light output. Another reason is that none of its component materials would require enrichment, as neither oxygen nor silicon have isotopes with high neutron-capture cross-sections.

LiF has the highest concentration of lithium atoms, but unfortunately its decay constant is far too long to be a viable LGB replacement. It may be possible that a dopant other than tungsten would result in shorter scintillation decay time, but such a combination is unknown to the authors.

CLYB is a known scintillator that has intrinsic neutron / gamma discrimination properties. Unfortunately, its lithium content is low (1 out of 10 atoms) relative to LGB (6 out of 19 atoms). Still, its high light output and long scintillation decay constant may make it a worthwhile LGB replacement . LTO has a high lithium content and high light output, but its scintillation decay time is too long to make it of use in a high-rate environment, as any detector using LTO would be limited to neutron capture rate of roughly 30 kHz. The very short decay constant of LTB would make its signal near indistinguishable from that of the plastic scintillator, or even Cherenkov light produced in the nonscintillating plastic detector.

%
%
\section{Applications}
\label{s:Applications}

Several applications are available for the detectors discussed in this work. One of the more immediate applications would be as a neutron scalar detector. Given that the continuing shortage in the \iso{3}{He} supply is not expected to diminish in the future, these detectors, with the improvements discussed in Sections~\ref{s:Optimizations} and \ref{ss:LGBAlternatives}, could be used as a replacement for \iso{3}{He}-based proportional counters. These detectors would have intrinsic efficiency for detecting fission neutrons in the 10-20\% range, while maintaining a strong gamma rejection capability of at least $10^{-8}$. Either the scintillating-plastic or nonscintillating-plastic detector would work for this application, although for a high gamma radiation environment, the nonscintillating-plastic detector would be preferable to reduce the overall trigger rate.

Another application would be an improved capture-gated spectrometer, as described in~\cite{Czirr2002}. The higher neutron capture rate afforded the various optimizations, combined with the stronger pulse shape analysis described in Section~\ref{ss:DAQchanges}, would allow for a more accurate, and higher-efficiency, spectrometer. Such a detector would be made of the scintillating plastic.

Finally, the scintillating-plastic detector could possibly be used for anti-neutrino detection. These detectors rely on a hydrogenous material to provide a target for inverse beta decays, and then look for a double signal coming from the prompt positron annihilation, followed by the neutron capture~\cite{Bowden2007}. Previous anti-neutrino detectors allow for some event selection based on energy deposition, but they largely do not utilize pulse shape in differentiating between neutron captures and gamma recoils. Using a scintillating-plastic detector of the kind described in this work would reduce background from both accidental gamma coincidences as well as multiple-neutron capture resulting from a single cosmic ray spallation event near the detector.

%
%
\section{Conclusions}
\label{s:Conclusions}

Lithium gadolinium borate crystals do not have any intrinsic neutron / gamma discrimination, but if small shards are embedded in a plastic matrix, the resulting signal can be used to differentiate neutron captures and gamma recoils with a high degree of confidence. The discrimination relies on accurate pulse shape identification, specifically between the fast-decaying signal associated with gamma recoils and the slow-decaying signal associated with neutron capture. Energy analysis also provides for particle discrimination, leading to gamma sensitivity on the order of $10^{-9}$.

The intrinsic neutron efficiency of the detectors studied in this work is on the order of 0.4-0.7\%, although they were not optimized for neutron detection. Four avenues exist for increasing neutron sensitivity, outlined in Table~\ref{tab:OptimizationTally}. These efficiencies are of course not orthogonal, although we expect that if all methods were employed, the performance of these detectors would surpass that of \iso{3}{He}-based detectors. From a quick sample survey, the most promising LGB alternatives are lithium yttrium borate and lithium glass. The gamma sensitivity could be kept low by decreasing the shard size, as well as using a more finely-tuned data acquisition system designed for this kind of composite detector. Even a factor of 15 improvement on the current neutron detection sensitivity would give these detectors a neutron efficiency comparable to moderated \iso{3}{He} tubes.

\begin{table}[t!!!]
\caption{\small{Optimizations to the LGB detectors, and their anticipated increase in fast neutron efficiency over the baseline performance.}}
\begin{center}
\begin{tabular}{|l|c|}
\hline
Method & Factor improvement over baseline \\
\hline
~2" polyethylene moderator & 10 \\
~Alternative crystal & 6 \\
~Increased total crystal content & 4 \\
~Smaller shards to limit self-shielding~~ & 1.5 \\
\hline
\end{tabular}
\end{center}
\label{tab:OptimizationTally}
\end{table}

Applications for these detectors include neutron counting (specifically \iso{3}{He}-based detector replacement), capture-gated neutron spectroscopy, and anti-neutrino detection.

This work performed under the auspices of the U.S. Department of Energy by Lawrence Livermore National Laboratory under Contract DE-AC52-07NA27344. LLNL-JRNL-499156.

%
%
\bibliography{apssamp}

\begin{thebibliography}{99}
\bibitem{LUXSim} LUXSim is not yet available to the public. Please contact the authors of this paper for additional information.
\end{thebibliography}

\end{document}